\documentclass[12pt]{article}

\usepackage{amsmath,amssymb,cite}
\usepackage{graphicx}

\makeatletter
 
  \@addtoreset{equation}{section}
 \makeatother

\newcommand{\be}{\begin{equation}}
\newcommand{\ee}{\end{equation}}
\newcommand{\bea}{\begin{eqnarray}}
\newcommand{\eea}{\end{eqnarray}}
\newcommand{\beann}{\begin{eqnarray*}}
\newcommand{\eeann}{\end{eqnarray*}}

\newcommand{\ba}{\begin{array}}
\newcommand{\ea}{\end{array}}

\newcommand{\tr}{\mbox{tr}}
\newcommand{\Tr}{\mathop{\rm Tr}}

\newcommand{\diag}{\mathop{\rm diag}\nolimits}

\newcommand{\del}{\partial}

\newcommand{\calZ}{{\cal Z}}

\newcommand{\av}[1]{\langle #1 \rangle}
\newcommand{\n}{\nonumber \\}

\def\XXint#1#2#3{{\setbox0=\hbox{$#1{#2#3}{\int}$} 
\vcenter{\hbox{$#2#3$}}\kern-.5\wd0}}

\allowdisplaybreaks

\begin{document}

\setlength{\oddsidemargin}{0cm}
\setlength{\baselineskip}{7mm}

\begin{titlepage}
\renewcommand{\thefootnote}{\fnsymbol{footnote}}
\begin{normalsize}
\begin{flushright}
\begin{tabular}{l}
KUNS-2251 \\
HRI/ST/1001
\end{tabular}
\end{flushright}
  \end{normalsize}

~~\\

\vspace*{0cm}
    \begin{Large}
       \begin{center}
         {A Novel Large-$N$ Reduction on $S^3$: \\
            Demonstration in Chern-Simons Theory}
       \end{center}
    \end{Large}
\vspace{0.7cm}

\begin{center}
Goro I{\sc shiki}$^{1)}$\footnote
            {
e-mail address : 
ishiki@post.kek.jp}, 
Shinji S{\sc himasaki}$^{2),3)}$\footnote
            {
e-mail address : 
shinji@gauge.scphys.kyoto-u.ac.jp}
    {\sc and}
Asato T{\sc suchiya}$^{4)}$\footnote
           {
e-mail address : 
satsuch@ipc.shizuoka.ac.jp}\\
      
\vspace{0.7cm}
                    
       $^{1)}$ {\it Center for Quantum Spacetime (CQUeST)
                    }\\
               {\it Sogang University, Seoul 121-742, Korea}\\
      \vspace{0.3cm}
      $^{2)}$ {\it Department of Physics, Kyoto University}\\
               {\it Kyoto, 606-8502, Japan}\\
      \vspace{0.3cm}
      $^{3)}$ {\it Harish-Chandra Research Institute}\\
               {\it Chhatnag Road, Jhusi,
                Allahabad 211019, India }\\
      \vspace{0.3cm}
      $^{4)}$ {\it Department of Physics, Shizuoka University}\\
               {\it 836 Ohya, Suruga-ku, Shizuoka 422-8529, Japan}
               
\end{center}

\vspace{0.7cm}

\begin{abstract}
\noindent
We show that the planar Chern-Simons (CS) theory
on $S^3$ can be described by its dimensionally reduced model.
This description of CS theory can be regarded as a novel 
large-$N$ reduction for gauge theories on $S^3$. 
We find that if one expands the reduced model around
a particular background consisting of multiple fuzzy spheres,
the reduced model becomes equivalent to CS theory on $S^3$ 
in the planar limit.
In fact, we show that the free energy and the vacuum expectation value of
unknot Wilson loop in CS theory are reproduced
by the reduced model in the large-$N$ limit.
\end{abstract}
\vfill
\end{titlepage}
\vfil\eject

\setcounter{footnote}{0}

\tableofcontents

\section{Introduction}
The large-$N$ reduction \cite{EK} asserts that large-$N$
gauge theories on flat space-times are equivalent 
to matrix models (reduced models) that are
obtained by dimensional reduction to lower dimensions (for further developments,
see \cite{Bhanot:1982sh,Parisi:1982gp,Gross:1982at,Das:1982ux,GonzalezArroyo:1982hz}).
It is not only conceptually relevant because 
it realizes emergent space-time in matrix models, but also practically relevant because it can give
a non-perturbative formulation of planar gauge theories as an alternative to
lattice gauge theories. 
It is well-known that the equivalence does not hold naively due to
the $U(1)^d$ symmetry breaking \cite{Bhanot:1982sh}. Some remedy is needed for the equivalence
to indeed hold (for recent studies, see \cite{Narayanan:2003fc,Kovtun:2007py,Unsal:2008ch,Bringoltz:2009kb,Poppitz:2009fm}).
However, as far as we know, there has been no remedy that manifestly preserves supersymmetry 
in  gauge theories.
Recently, the large-$N$ reduction was generalized 
to theories on $S^3$ in \cite{Ishii:2008ib}
(for earlier discussions, see \cite{Ishiki:2006yr,Ishii:2007ex,Ishii:2008tm}).
Here $S^3$ is constructed
as a nontrivial $U(1)$ fiber bundle over $S^2$ by expanding matrix models
around a particular background corresponding to a sequence of fuzzy spheres.
This novel large-$N$ reduction is important from both of the above two viewpoints.
It would give hints to the problem of describing curved space-time \cite{Hanada:2005vr} in the matrix models \cite{BFSS,IKKT,DVV}.
Unlike on flat space-times, the equivalence would hold on curved space-times without any remedy 
since the theories are massive due to the curvature of $S^3$.
The novel large-$N$ reduction can, therefore, give a non-perturbative regularization of gauge theories 
that respects both supersymmetry and the gauge symmetry\footnote{For recent developments in lattice approach to
supersymmetric gauge theories, see \cite{Catterall:2009kg} and references therein.}. 

The novel large-$N$ reduction has been studied mainly for two cases so far.
One is ${\cal N}=4$ SYM 
on $R\times S^3$ 
\cite{Ishii:2008ib,Ishiki:2008te,Ishiki:2009sg,Kitazawa:2008mx}.
The reduced model of this theory takes the form of  
the plane wave matrix model \cite{Berenstein:2002jq}.
The large-$N$ reduction provides a non-perturbative formulation 
of ${\cal N}=4$ SYM, which respects sixteen supersymmetries 
and the gauge symmetry
so that no fine tuning of the parameters is required.
Thus the formulation gives a feasible way to analyze 
the strongly coupled regime of ${\cal N}=4$ SYM, 
for instance, by putting it on a computer 
through the methods proposed in
\cite{Hanada,Anagnostopoulos,Catterall:2008yz}, 
and therefore enables us to perform new nontrivial tests of 
the AdS/CFT correspondence \cite{Maldacena,GKP,Witten}.

The other is Chern-Simons (CS) theory 
on $S^3$, which has been
exactly solved \cite{Witten:1988hf}.
In this case, the equivalence between the reduced model 
and the original theory
can be verified explicitly, which was briefly reported 
in \cite{Ishiki:2009vr}.
In this paper, we present the verification 
of the equivalence in detail.
We also discuss the large-$N$ reduction for CS theory on $S^3/Z_q$.
Our result is an explicit demonstration
of the novel large-$N$ reduction.
CS theory on $S^3$ is a topological field theory 
associated with the knot theory and therefore interesting in its own right.
Our formulation gives a new regularization method for CS theory on $S^3$.
CS theory on $S^3$ is also interpreted as open 
topological A strings on $T^{\ast}S^3$ in
the presence of $N$ D-branes wrapping $S^3$. Topological strings 
capture some aspects of more realistic string, so that
our study is expected to give some insights into
the matrix models as a non-perturbative formulation of superstring.

More recently, it was shown in \cite{Kawai:2009vb} 
that the large-$N$ reduction holds on general group manifolds.
In the case of $S^3\simeq SU(2)$, the background 
taken in the large-$N$ reduction in \cite{Kawai:2009vb} 
is different from that in \cite{Ishii:2008ib}.
It is desirable to explicitly demonstrate
this different type 
of the large-$N$ reduction, for instance, in the case of CS theory on $S^3$. 
We leave this for a future study.

This paper is organized as follows. 
In section 2, we briefly review CS theory on $S^3$.
In section 3, we summarize 
the relationships among CS theory on $S^3$, 2d YM on $S^2$ and 
the matrix model which were obtained in
\cite{Ishii:2007sy,Ishiki:2008vf}.
In section 4, we argue how CS theory on $S^3$ is realized in the matrix 
model as a novel large-$N$ reduction.
In section 5, we show the equivalence between the theory around a 
particular background of the matrix model
and CS theory on $S^3$.
In section 6, by using the Monte Carlo simulation, 
we give a check of this equivalence.
Section 7 is devoted to conclusion and discussion. 
In appendices, some details are gathered.

\section{Chern-Simons theory on $S^3$}
In this section, we review some known facts about CS theory on $S^3$ needed for this paper
\cite{Marino:2004uf,Marino:2004eq}.
We consider CS theory on $S^3$ with the gauge group $U(N)$, whose action is given by
\begin{align}
S_{CS}=\frac{k}{4\pi}\int_{S^3}\mbox{Tr}\left(A\wedge dA+\frac{2}{3}A\wedge A\wedge A\right),
\label{Chern-Simons theory on S^3}
\end{align}
where $k$ must be an integer.
The partition function
\begin{align}
{\cal Z}_{CS}=\int {\cal D}A\; e^{iS_{CS}}
\label{partition function of CS}
\end{align}
defines a topological invariant of the manifold $S^3$, which also depends on a choice of framing.
Given an oriented knot ${\cal K}$ in $S^3$, one can consider the Wilson loop in an irreducible representation
$R$ of $U(N)$
\begin{align}
W_R({\cal K})=\mbox{Tr}_R \left(P\exp \oint_{{\cal K}}A \right).
\end{align}
The expectation value of the Wilson loop 
\begin{align}
\langle W_R({\cal K}) \rangle_{CS}=\frac{1}{{\cal Z}_{CS}}
\int {\cal D}A\; W_R({\cal K})\; e^{iS_{CS}}
\end{align}
defines a topological invariant
of ${\cal K}$ depending on a choice of framing as well. 
In this paper, we are concerned with the Wilson loop for the unknot, which we denote by $W_R(\mbox{unknot})$,
and mainly consider the Wilson loop for the unknot in the 
fundamental representation, which we denote
by $W_{\square}(\mbox{unknot})$.

$S^3$ is obtained by gluing two solid tori along their boundaries with the transformation
\begin{align}
K=T^mST^n,
\end{align}
where $S$ and $T$ are generators of $SL(2,Z)$, and $m$ and $n$ are arbitrary integers.
$T$ has the following representation in the Hilbert space ${\cal H}(T^2)$, which is obtained by doing canonical quantization of CS theory on $R\times T^2$:
\begin{align}
&\langle R|T|R'\rangle=\delta_{RR'}T_R, \\
&T_R=\exp \left[\frac{2\pi i}{k+N}\left(\Lambda\cdot\rho +\frac{1}{2}\Lambda^2-\frac{1}{24}k(N^2-1)\right)\right],
\end{align}
where $|R\rangle$ denotes a state associated to the highest weight $\Lambda$, and
$\rho$ is the Weyl vector of $SU(N)$.
The partition function is given by
\begin{align}
{\cal Z}_{CS}= \langle 0|K|0\rangle,
\label{partition function by surgery}
\end{align}
where $0$ stands for the trivial representation, while the expectation value of the Wilson loop for the unknot is given by
\begin{align}
\langle W_R(\mbox{unknot})\rangle_{CS}= \frac{\langle R|K|0\rangle}{\langle 0|K|0\rangle}.
\label{Wilson loop by surgery}
\end{align}

The canonical framing corresponds to $m=n=0$.
In the following, we first present the exact results for the partition function and the expectation value
of the Wilson loop in the canonical framing.
The partition function (\ref{partition function of CS}) was computed exactly:
\begin{align}
{\cal Z}_{CS}^{m=n=0}&=\langle0|S|0\rangle
=\frac{1}{(k+N)^{N/2}}\sum_{w\in W}\varepsilon(w) \exp \left(-\frac{2\pi i}{k+N}\rho\cdot w(\rho)\right),
\label{result for the partition function}
\end{align}
where the sum over $w$ is a sum over the elements of the Weyl group ${\cal W}$ of $U(N)$, and $\varepsilon(w)$ is 
the signature of $w$. The expectation value of the Wilson loop
for the unknot was also computed exactly:
\begin{align}
\langle W_R(\mbox{unknot}) \rangle_{CS}^{m=n=0} = \frac{\langle R|S|0\rangle}{\langle 0|S|0\rangle}
= \frac{\sum_{w\in W}\varepsilon(w) e^{-\frac{2\pi i}{k+N}\rho\cdot w(\Lambda+\rho)}}
{\sum_{w\in W}\varepsilon(w) e^{-\frac{2\pi i}{k+N}\rho\cdot w(\rho)}}.
\label{result for the Wilson loop}
\end{align}

One can obtain different expressions for (\ref{result for the partition function}) and 
(\ref{result for the Wilson loop}) by using Weyl's character formula
\begin{align}
\mbox{ch}_R(u)=\sum_{\mu\in M_R}e^{\mu\cdot u}
=\frac{\sum_{w\in W}\varepsilon(w)e^{w(\Lambda+\rho)\cdot u}}{\prod_{\alpha>0}2\sinh\frac{\alpha\cdot u}{2}},
\label{character formula}
\end{align}
and Weyl's denominator formula
\begin{align}
\sum_{w\in W}\varepsilon(w) e^{w(\rho)\cdot u}=\prod_{\alpha>0}2\sinh \frac{\alpha\cdot u}{2},
\label{denominator formula}
\end{align}
where $\alpha >0$ are positive roots, and $M_R$ is the set of weights associated to the representation $R$.
By applying (\ref{denominator formula}) to (\ref{result for the partition function}), one obtains
\begin{align}
{\cal Z}_{CS}^{m=n=0}
=\frac{1}{(k+N)^{N/2}}\prod_{\alpha>0}2\sinh \left(\frac{2\pi i}{k+N}\frac{\alpha\cdot\rho}{2}\right),
\end{align}
while by applying (\ref{character formula}) and (\ref{denominator formula}) to (\ref{result for the Wilson loop})
\begin{align}
\langle W_R(\mbox{unknot}) \rangle_{CS}^{m=n=0} 
=\mbox{ch}_R\left(-\frac{2\pi i}{k+N}\rho\right) 
=\prod_{\alpha>0}
\frac{\sinh \left(\frac{2\pi i}{k+N}\frac{\alpha\cdot(\Lambda+\rho)}{2}\right)}
{\sinh \left(\frac{2\pi i}{k+N}\frac{\alpha\cdot\rho}{2}\right)}.
\end{align}
For the fundamental representation,
it reduces to
\begin{align}
\langle W_{\square}(\mbox{unknot}) \rangle_{CS}^{m=n=0}
=\frac{\sinh\frac{\pi i N}{k+N}}{\sinh\frac{\pi i}{k+N}}.
\end{align}

Furthermore, by using (\ref{denominator formula}), 
one can show that (\ref{result for the partition function}) is equivalent to
an integral over $N$ variables which takes the form analogous to a partition function of a matrix model \cite{Tierz:2002jj,Dolivet:2006ii,Marino:2004uf}:
\begin{align}
{\cal Z}_{CS}^{m=n=0}
={\cal Z}_{CSM}=\frac{e^{-\frac{g_s}{12}N(N^2-1)}}{N!}\int\prod_{i=1}^N\frac{d\beta_i}{2\pi}e^{-\frac{1}{2g_s}\sum_i\beta_i^2}\prod_{i<j}\left(2\sinh\frac{\beta_i-\beta_j}{2}\right)^2,
\label{CSM}
\end{align}
where we have introduced
\begin{align}
g_s=\frac{2\pi i}{k+N},
\end{align}
which is identified with the string coupling in the context of topological strings.
We call the statistical model defined by (\ref{CSM}) the Chern-Simons matrix model (CS matrix model).
Correspondingly, one can find a relation
\begin{align}
\langle W_R(\mbox{unknot}) \rangle_{CS}^{m=n=0}
=e^{-g_s(\rho\cdot\Lambda+\frac{1}{2}\Lambda^2)}\langle \mbox{ch}_R(\beta) \rangle_{CSM}
=e^{-g_s(\rho\cdot\Lambda+\frac{1}{2}\Lambda^2)}\left\langle \sum_{\mu\in M_R}e^{\mu\cdot\beta} \right\rangle_{CSM}
\end{align}
where
\begin{align}
\langle \cdots \rangle_{CSM}
=\frac{1}{Z_{CSM}}\frac{e^{-\frac{1}{12}N(N^2-1)}}{N!}\int\prod_{i=1}^N\frac{d\beta_i}{2\pi} (\cdots) \:
e^{-\frac{1}{2g_s}\sum_i\beta_i^2}\prod_{i<j}\left(2\sinh\frac{\beta_i-\beta_j}{2}\right)^2 .
\end{align}

Finally we consider
the framing with $m=n=1$, which appears in the direct evaluation of the partition function
of CS theory on $S^3$ in \cite{Blau:2006gh}.  
One can see from (\ref{partition function by surgery}) that the 
partition function in this framing is related to 
the one in the canonical framing as
\begin{align}
{\cal Z}_{CS}^{m=n=1}&=T_0^2{\cal Z}_{CS}^{m=n=0}\nonumber\\
&=\frac{e^{-\frac{\pi i}{6}(N^2-1)+\frac{g_s}{12}N(N^2-1)}}{(k+N)^{N/2}}\prod_{\alpha>0}2\sinh \left(\frac{2\pi i}{k+N}\frac{\alpha\cdot\rho}{2}\right) \nonumber\\
&=\frac{e^{-\frac{\pi i}{6}(N^2-1)}}{N!}\int\prod_{i=1}^N\frac{d\beta_i}{2\pi}e^{-\frac{1}{2g_s}\sum_i\beta_i^2}\prod_{i<j}\left(2\sinh\frac{\beta_i-\beta_j}{2}\right)^2.
\label{partition function in non-canonical framing} 
\end{align}
One can also see from (\ref{Wilson loop by surgery}) that
the expectation value of the Wilson loop is related to the one in the canonical framing as
\begin{align}
\langle W_R(\mbox{unknot}) \rangle_{CS}^{m=n=1}
=\frac{T_R}{T_0}\langle W_R(\mbox{unknot}) \rangle_{CS}^{m=n=0}
=\langle \mbox{ch}_R(\beta) \rangle_{CSM}=\left\langle \sum_{\mu\in M_R}e^{\mu\cdot\beta} \right\rangle_{CSM}.
\label{Wilson loop in non-canonical framing} 
\end{align}
In particular, because $\Lambda\cdot\rho+\frac{1}{2}\Lambda^2=\frac{N^2-1}{2N}$ for the fundamental representation,
one obtains
\begin{align}
\langle W_{\square}(\mbox{unknot}) \rangle_{CS}^{m=n=1}
=e^{\frac{g_s}{2}\left(N-\frac{1}{N}\right)}\frac{\sinh\frac{g_sN}{2}}{\sinh\frac{g_s}{2}}.
\label{WL in noncanonical framing}
\end{align}
Similarly, the perturbative part of the free energy 
is evaluated in this framing as
\begin{align}
{\cal F}_{CS}=\frac{g_s}{12}N(N^2-1)+\sum_{j=1}^{N}(N-j)\sum_{n=1}^{\infty}
\log \left(1+\frac{j^2g_s^2}{4\pi^2n^2}\right).
\label{free energy in CS}
\end{align}

In the following sections, we will show that the reduced model of CS theory on $S^3$ reproduces
the results for the original theory in the framing corresponding $m=n=1$.

\section{CS theory on $S^3$, 2d YM on $S^2$ and a matrix model}
In this section, we review part of the results in 
\cite{Ishii:2007sy,Ishiki:2008vf}, which we use in this paper.
In section 3.1, we dimensionally reduce CS theory on $S^3$ to 2d YM on $S^2$ and a matrix model.
In section 3.2, we describe a classical equivalence between
the theory around each multiple monopole background of 2d YM on $S^2$
and the theory around a certain multiple fuzzy sphere background of the matrix model.
In section 3.3, we exactly perform the integration in the matrix model. 
In section 3.4, using the result in section 3.3, we show that the 
equivalence in section 3.2 also holds at quantum level.

\subsection{Dimensional reductions}
In order to dimensionally reduce CS theory on $S^3$, 
we expand the gauge field in terms of the right-invariant 1-form defined in (\ref{right-invariant 1-form}) as
\begin{align}
A=iX_iE^i.
\end{align}
We choose an $SO(4)$-isometric metric for $S^3$ (\ref{metric of S^3})
fixing the radius of $S^3$ to $2/\mu$.
Then, we rewrite (\ref{Chern-Simons theory on S^3}) as
\begin{align}
S_{CS}=-\frac{k}{4\pi}\int \frac{d\Omega_3}{(\mu/2)^3}
\mbox{Tr}\left(i\mu\epsilon^{ijk}X_i{\cal L}_jX_k+\mu X_i^2+\frac{2i}{3}\epsilon^{ijk}X_iX_jX_k\right),
\end{align}
where ${\cal L}_i$ is the Killing vector dual to $E^i$ defined in (\ref{definition of Killing vector}),
and we have used the Maurer-Cartan equation (\ref{Maurer-Cartan}).

As we explain in appendix \ref{$S^3$ and $S^2$}, 
we regard $S^3$ as a $U(1)$ bundle over $S^2$. 
Here the fiber direction is parametrized by $y$.
By dropping the $y-$derivatives, we obtain a gauge theory on $S^2$:
\begin{align}
S_{BF}=
\frac{\mu}{2g_{YM}^2}\int \frac{d\Omega_2}{\mu^2}
\mbox{Tr}\left(i\mu\epsilon^{ijk}X_iL_jX_k+\mu X_i^2+\frac{2i}{3}\epsilon^{ijk}X_iX_jX_k\right),
\label{BF with mass term on S^2}
\end{align}
where $g_{YM}^2=-\frac{\mu^2}{2k}$\footnote{While $k$ in (\ref{Chern-Simons theory on S^3}) must be an integer, 
such a restriction is not imposed on $k$ in (\ref{BF with mass term on S^2}).},
the radius of $S^2$ is $1/\mu$, and
$L_i$ are the angular momentum operators 
on $S^2$ given in (\ref{monopole angular momentum}) with $q=0$.
$g_{YM}$ will be identified with the coupling constant 
of 2d YM on $S^2$ below.
First, we see that (\ref{BF with mass term on S^2}) is the BF theory with a mass term on $S^2$.
We expand $X_i$ as \cite{Kitazawa:2002xj,Ishii:2008tm}
\begin{align}
\vec{X}=\mu\left(\chi\vec{e}_r+a_{\theta}\vec{e}_{\varphi}-\frac{1}{\sin\theta}a_{\varphi}\vec{e}_{\theta}\right),
\end{align}
where $\vec{e}_r=(\sin\theta\cos\varphi,\sin\theta\sin\varphi,\cos\theta)$ and
$\vec{e}_{\theta}=\frac{\partial\vec{e}_r}{\partial\theta}$, 
$\vec{e}_{\varphi}=\frac{1}{\sin\theta}\frac{\partial\vec{e}_r}{\partial\varphi}$.
$a_{\mu}$ are the gauge field on $S^2$, and $\chi$ is a scalar field on $S^2$ which comes from
the component of the fiber direction of the gauge field on $S^3$.
Indeed, we can rewrite (\ref{BF with mass term on S^2}) as
\begin{align}
S_{BF}=
\frac{\mu^4}{2g_{YM}^2}
\int\frac{d\Omega_2}{\mu^2}
\mbox{Tr}\left(\chi\epsilon^{\mu\nu} f_{\mu\nu}-\chi^2\right),
\label{BF with mass term on S^2 2}
\end{align}
where $f_{\mu\nu}=\del_\mu a_\nu-\del_\nu a_\mu+i[a_\mu,a_\nu]$ is the field strength.
This is the BF theory with a mass term:
the first term is the BF term and the second term is a mass term.
Next, we integrate $\chi$ out in (\ref{BF with mass term on S^2 2}) to obtain
2d YM on $S^2$:
\begin{align}
S_{YM}=\frac{\mu^4}{g_{YM}^2}\int \frac{d\Omega_2}{\mu^2}
\mbox{Tr}\left(\frac{1}{4}f^{\mu\nu}f_{\mu\nu}\right).
\label{2d YM}
\end{align}

Furthermore, by dropping all the derivatives in (\ref{BF with mass term on S^2}) and rescale $X_i$ as 
$X_i\rightarrow \mu X_i$, we obtain a three-matrix model:
\begin{align}
S_{m}=-\frac{1}{g_m^2}\mbox{Tr}\left(X_i^2+\frac{i}{3}\epsilon^{ijk}X_i[X_j,X_k]\right),
\label{N=1^* matrix model}
\end{align}
where 
$1/g_m^2=-2\pi \mu^2/g_{YM}^2$.
In the sense of the Dijkgraaf-Vafa theory \cite{Dijkgraaf:2002fc}, this matrix model is regarded as
a mass deformed superpotential of ${\cal N}=4$ SYM, which gives the so-called ${\cal N}=1^*$ theory. 

\subsection{Classical equivalence between 2d YM on $S^2$ and the matrix model}
The matrix model (\ref{N=1^* matrix model}) with the matrix size $M\times M$
possesses the following classical solution,
\begin{align}
\hat{X}_i=L_i=\bigoplus_s L^{[j_s]}_i \otimes 1_{N_s}
\label{matrix background}
\end{align}
where $L^{[j_s]}_i$ are the spin $j_s$ representation of the $SU(2)$ generators 
obeying $[L^{[j_s]}_i,L^{[j_s]}_j]=i\epsilon_{ijk}L^{[j_s]}_k$, and the relation $\sum_{s}(2j_s+1)N_s=M$ is 
satisfied. $s$ runs over some integers, 
but its range is not specified here.
It will be specified later as 
$-\Lambda/2 \leq s \leq \Lambda/2$ with $\Lambda$ an even positive integer. 

We put $2j_s+1=\Omega+n_s$ with $\Omega$ and $n_s$ integers and take the limit in which 
\begin{align}
\Omega\rightarrow\infty \;\;\; \mbox{with} \;\;\; 
\frac{\Omega}{g_m^2}
=-\frac{8\pi^2}{g_{YM}^2A}=\mbox{fixed},
\label{limit}
\end{align}
where $A=4\pi/\mu^2$ is the area of $S^2$.
Then,  it was shown classically in \cite{Ishii:2007sy} that
the theory around (\ref{matrix background}) is equivalent to the theory around
the following classical solution of (\ref{BF with mass term on S^2}) with the gauge group $U(K)$,
\begin{align}
\mu L_i+\hat{X}_i=\mu\mbox{diag}(\cdots,\underbrace{L^{(q_s)}_i,\cdots,L^{(q_s)}_i}_{N_s}, \cdots),
\label{monopole solution 1}
\end{align}
where $q_s=n_s/2$, and $\sum_sN_s=K$ is satisfied. $L^{(q_s)}_i$ are the angular momentum operators in the presence of a monopole
with the monopole charge $q_s$, which are given in (\ref{monopole angular momentum}). 
This theory can also be viewed as the theory around the following classical solution
of (\ref{BF with mass term on S^2 2}),
\begin{align}
&\hat{\chi}=-\mbox{diag}(\cdots,\underbrace{q_s,\cdots,q_s}_{N_s}, \cdots),\nonumber\\
&\hat{a}_{\theta}=0, \nonumber\\
&\hat{a}_{\varphi}=(\cos\theta \mp 1) \hat{\chi},
\label{monopole solution}
\end{align}
where the upper sign is taken in the region $0\leq \theta < \pi$ and the lower sign
in the region $0 < \theta \leq \pi$, and $\hat{a}_{\theta}$ and $\hat{a}_{\varphi}$ represent
the monopole configuration. Equivalently, this theory can be viewed as the theory around
the multiple monopole background with the background for the gauge field given in (\ref{monopole solution}) of 2d YM on $S^2$ (\ref{2d YM}).
As reviewed in the following two subsections, the above equivalence is extended to quantum level \cite{Ishiki:2008vf}.

\subsection{Exact integration of the matrix model}
First, we rewrite the matrix model (\ref{N=1^* matrix model}) in terms of the following matrices;
\begin{align}
Z&=X_1+iX_2, \nonumber\\
\Phi&=X_3,
\end{align}
where $Z$ is an $M\times M$ complex matrix and $\Phi$ is an $M\times M$ hermitian matrix. 
The partition function is defined by
\begin{align}
\calZ_{m}&=\frac{1}{{\rm vol}(U(M))} \int dZ dZ^\dagger d\Phi e^{iS_{m}}, \\
S_{m}&=-\frac{1}{g_m^2}\tr \left\{ Z[\Phi,Z^\dagger]+ (1-i\epsilon)ZZ^\dagger+(1-i\epsilon)\Phi^2
\right\}.
\end{align}
Here, in order to make the path integral well-defined and converge, 
we introduce $i$ in the exponential and $-i\epsilon$ in the action. 
Taking the gauge in which $\Phi$ is diagonal as $\Phi=\diag(\phi_1,\phi_2,\cdots, \phi_M)$ and
integrating $Z$ and $Z^\dagger$ out, then we obtain
\begin{align}
\calZ_{m}=\frac{1}{(2\pi)^M M!}\left(\frac{g_m^2\pi}{i}\right)^{M^2}
  \int \prod_i d\phi_i
 \prod_{i\neq j}\frac{\phi_i-\phi_j}{\phi_i-\phi_j+1-i\epsilon}
 e^{-\frac{i}{g_m^2}(1-i\epsilon)\sum_i\phi_i^2}, \label{calZ_{m} in eigenvalues}
\end{align}
where $\prod_{i\neq j}(\phi_i-\phi_j)$ comes from the diagonalization of $\Phi$, which is the square
of the Vandermonde determinant. $(2\pi)^M$ is the volume of $U(1)^M$ and
$M!$ is the volume of the Weyl group of $U(M)$.

Next let us concentrate on the factor in the integrand in (\ref{calZ_{m} in eigenvalues}),
\begin{align}
\prod_{i\neq j}\frac{\phi_i-\phi_j}{\phi_i-\phi_j+1-i\epsilon}.
\label{prod factor in calZ_{m}}
\end{align}
For this factor, we use the following formula,
\begin{align}
\frac{1}{x-i\epsilon}=P.V. \frac{1}{x}+i\pi\delta(x),
\end{align}
where ``$P.V.$'' denotes the principal value.
Then we obtain 
\begin{align}
&\prod_{i\neq j}\frac{\phi_i-\phi_j}{\phi_i-\phi_j+1-i\epsilon} \nonumber\\
&=\prod_{i<j}\left[
P.V. \frac{(\phi_i-\phi_j)(\phi_j-\phi_i)}{(\phi_i-\phi_j+1)(\phi_j-\phi_i+1)}
 +\left(-\frac{i\pi}{2}\right)\delta(\phi_i-\phi_j+1)
 +\left(-\frac{i\pi}{2}\right)\delta(\phi_j-\phi_i+1)
 \right]. \label{expanded prod factor}
\end{align}
It is easily seen that (\ref{expanded prod factor}) is written in the sum of terms containing
\begin{align}
&\left(-\frac{i\pi}{2}\right)^{2j_1}
\delta(\phi_1^{(1)}-\phi_2^{(1)}-1)\delta(\phi_2^{(1)}-\phi_3^{(1)}-1)\cdots
\delta(\phi_{2j_1}^{(1)}-\phi_{2j_1+1}^{(1)}-1) \nonumber\\
&\times
\left(-\frac{i\pi}{2}\right)^{2j_2} 
\delta(\phi_1^{(2)}-\phi_2^{(2)}-1)\delta(\phi_2^{(2)}-\phi_3^{(2)}-1)\cdots
\delta(\phi_{2j_2}^{(2)}-\phi_{2j_2+1}^{(2)}-1) \nonumber\\
&\times \cdots \nonumber\\
&\times 
\left(-\frac{i\pi}{2}\right)^{2j_K}
\delta(\phi_1^{(K)}-\phi_2^{(K)}-1)\delta(\phi_2^{(K)}-\phi_3^{(K)}-1)\cdots 
\delta(\phi_{2j_K}^{(K)}-\phi_{2j_K+1}^{(K)}-1),
\label{product of delta functions}
\end{align}
where $K$ can take the value $1,2,\cdots, M$ and we have reordered and relabeled the eigenvalues of $\Phi$ as
\begin{align}
\Phi=\diag(\phi_1^{(1)},\cdots, \phi_{2j_1+1}^{(1)},
\phi_1^{(2)},\cdots,\phi_{2j_2+1}^{(2)},\cdots,
\phi_1^{(K)},\cdots, \phi_{2j_K+1}^{(K)}).
\end{align}
The size of the matrices, $M$, satisfies $M=\sum_{I=1}^{K}(2j_I+1)$. $\phi_i^{(I)}$ represents the $i$-th
component of the $I$-th block.
Thus (\ref{expanded prod factor}) is decomposed into terms, each of which is characterized by
an $M$-dimensional reducible representation of the $SU(2)$ algebra consisting of $K$ blocks (irreducible
representations), with a $U(1)$ degree of freedom in each block.
We label the reducible representations by ``$r$''\footnote{
Namely,  ``$r$'' specify the number of irreducible representations of $SU(2)$, $K$, and the dimensions of the
irreducible representations,  $2j_I+1\; (I=1,\cdots K)$, which satisfy the relation $M=\sum_{I=1}^{K}(2j_I+1)$.} and denote the $U(1)$ part in the $I$-th block by
$a_I$, putting $a_I\equiv \phi_{2j_I+1}^{(I)}+j_I$.
Then, we obtain
\begin{align}
\calZ_{m}
 &=\sum_r\frac{{\cal N}_r}{(2\pi)^M} (-i\pi)^{M-N} \left(\frac{g_m^2\pi}{i}\right)^{M^2}
 \prod_{I=1}^{K}\frac{1}{2j_I+1} P.V. \int \prod_{I=1}^{K}da_I \nonumber\\
 &\qquad \qquad \times
 \prod_{I\neq J}\prod_{m_I=-j_I}^{j_I}\prod_{m_J=-j_J}^{j_J}
 \frac{a_I+m_I-a_J-m_J}{a_I+m_I-a_J-m_J+1} 
 e^{-\frac{i}{g_m^2}(1-i\epsilon)\sum_{I=1}^{K}\sum_{m_I=-j_I}^{j_I}(a_I+m_I)^2},
\label{decomposed form 1}
\end{align}
where 
\begin{align}
{\cal N}_r\equiv \prod \frac{1}{(\text{$\sharp$ of blocks with the same length})!}.
\label{N_r}
\end{align}

After some calculations, we find that (\ref{decomposed form 1}) results in 
\begin{align}
 \calZ_{m}
 &=\sum_{r}\frac{{\cal N}_r}{(2\pi)^M} (-i\pi)^{M-N}\left(\frac{g_m^2\pi}{i}\right)^{M^2}
 \prod_{s=1}^{N}\frac{1}{2j_s+1}  e^{-\frac{i}{3g_m^2}(1-i\epsilon)\sum_{I=1}^{K}j_I(j_I+1)(2j_I+1) }\nonumber\\
 &\qquad \times P.V. \int \prod_{I=1}^{K}da_I
 \prod_{I<J}\frac{(a_I-a_J)^2-(j_I-j_J)^2}{(a_I-a_J)^2-(j_I+j_J+1)^2}
 e^{-\frac{i}{g_m^2}(1-i\epsilon)\sum_{I=1}^{K}  (2j_I+1)a_I^2}
\label{result MM partition function}
\end{align}
Thus the partition function of the matrix model is decomposed into sectors 
characterized by
the (reducible) representation of the $SU(2)$ algebra. 
Note that $P.V.$ prevents the integration over the $U(1)$ part $a_I$ from mixing the sectors.

\subsection{2d YM on $S^2$ from the matrix model}
In this subsection, we show that the partition function of YM on $S^2$ is obtained from that of
the matrix model \cite{Ishiki:2008vf}. 
As we will see, the sector which consists of $K$ irreducible representations in the matrix model
corresponds to $SU(K)$ 2d YM on $S^2$.
In the $K$ block sector, we group the irreducible representations with the same dimension and label the groups
by $s$, such that $j_s\neq j_t$ for $s\neq t$. We denote the multiplicity of the $s$-th group by $N_s$. 
Then, the relations 
$K=\sum_{s}N_s$ and $M=\sum_{s}N_s(2j_s+1)$ are satisfied.
We also  denote the $i$-th $U(1)$ part in the $s$-th group by $a_{si}$, 
where $i=1,\cdots, N_s$. Note that in this notation ${\cal N}_r$ in (\ref{N_r}) equals $1/(\prod_s N_s!)$.
 
Since the partition function of the matrix model (\ref{result MM partition function})
is completely decomposed into sectors without overlap,
we can extract the sectors consisting of $K$ blocks. 
In the $K$-block sector, due to the factor $\exp(-\frac{i}{3g_m^2}(1-i\epsilon)\sum_{I=1}^{K}j_I(j_I+1)(2j_I+1))$, 
configurations of almost equal size blocks are dominant
in the limit in which $M/K=\Omega\rightarrow \infty,\; g_m^2/\Omega=\text{fixed}$. 
So, we consider configurations around the dominated configuration by setting
\begin{align}
2j_s+1=\Omega+n_s,
\end{align}
where $|n_s|\ll\Omega$. 
Substituting this into the $K$-block sector in (\ref{result MM partition function})
and taking the limit (\ref{limit}), we obtain
\begin{align}
 {\cal Z}_{K\text{-block}}&=C\sum_{\{n_s,N_s\}}\int \prod_{s}\prod_{i=1}^{N_s}da_{si}
\prod_{s\leq t}\prod_{i=1}^{N_s}\prod_{j=1}^{N_t}
\left\{(a_{si}-a_{tj})^2-\frac{1}{4}(n_s-n_t)^2\right\} \nonumber\\
& \hspace{70mm} \times
e^{-i\frac{8\pi^2}{g_{YM}^2A}\sum_{s}\sum_{i=1}^{N_s}\left({a_{si}}^{2}+\frac{n_s^2}{4}\right)}.
\end{align}
Here we have absorbed overall irrelevant constants and divergences into a renormalized constant $C$.
We have also replaced the denominator in the integrand in (\ref{result MM partition function}) by
constant ($\Omega^2$)
and taken the integral regions of $a_{si}$ to $(-\infty,\infty)$,
since the exponential factor in (\ref{result MM partition function}) oscillates 
rapidly around $|a_{si}|\gtrsim \Omega$
in the limit where $\epsilon\rightarrow 0$ and $\Omega\rightarrow \infty$,
so that the integral dominates  around $|a_{si}|\ll \Omega$.

Finally, by rescaling $a_{si}$ by $y_{si}\equiv 2a_{si}$, and making an
analytical continuation $g_{YM}^2\rightarrow -ig_{YM}^2$, we obtain
\begin{align}
{\cal Z}_K&=C'\sum_{\{n_s,N_s\}}\int\prod_{s}\prod_{i=1}^{N_s}dy_{si}
\prod_{s\leq t}\prod_{i=1}^{N_s}\prod_{j=1}^{N_t}
\left\{(y_{si}-y_{tj})^2-(n_s-n_t)^2\right\} \nonumber\\
&\hspace{100mm} \times
e^{-\frac{2\pi^2}{g_{YM}^2A}\sum_{s}\sum_{i=1}^{N_s}(y_{si}^2+n_s^2)},
\label{2dYM mp}
\end{align}
where irrelevant constants are again absorbed into a constant $C'$. ${\cal Z}_K$  exactly agrees with the
partition function of $U(K)$ 2d YM on $S^2$
\cite{Migdal:1975zg,Witten:1992xu,Minahan:1993tp,Gross:1994mr,Blau:1993hj}.\footnote{In fact, only the $SU(K)$ part agrees, and
the $U(1)$ part does not (see ref.\cite{Ishiki:2008vf}). 
However, this difference does not matter in the following arguments.}
In YM on $S^2$, $n_s$ are identified with the monopole
charges since the $n_s$ dependent term in the exponential coincides with the contribution of the classical
solution (\ref{monopole solution}) to the action of 2d YM. This fact is consistent with the classical equivalence reviewed in section 3.2.

\section{CS theory on $S^3$ from the matrix model}
\label{CS theory on $S^3$ from the matrix model}
In this section, we provide a prescription for realization of  
CS theory on $S^3$ in 2d YM on $S^2$ and in the matrix model.
We also generalize the prescription to CS theory on $S^3/Z_q$.
In section 5, we prove the large-$N$ equivalence between CS theory and 
the matrix model based on this prescription.

CS theory on $S^3$ is realized from 2d YM on $S^2$
through the large-$N$ reduction 
for the case of compact space proposed in \cite{Ishii:2008ib}
as follows.
Note that the Kaluza-Klein (KK) momenta along the fiber ($S^1$)
direction in $S^3$ are regarded as the monopole charges on $S^2$.
Hence, if one considers the particular monopole solution 
in 2d YM on $S^2$ 
such that the spectrum of the monopole charges reproduces 
the spectrum of the KK momenta along $S^1$, 
the theory on $S^2$ expanded around such monopole solution
is equivalent to CS theory on $S^3$.
More precisely, in order to realize CS theory on $S^3$,
we first put $n_s=s$ and $N_s=N$ in (\ref{monopole solution 1})
(or in (\ref{monopole solution})) and 
then make $s$ run from $-\Lambda/2$ to 
$\Lambda/2$, where $\Lambda$ is 
a positive even integer.
Finally we take the following limit,
\begin{align}
\Lambda\rightarrow\infty, \;\;\; N\rightarrow \infty, \;\;\; 
-\frac{g_{YM}^2AN}{2\pi}=\frac{N}{k+N}
=\mbox{fixed}.
\label{limit 1}
\end{align}
Then, we see that
2d YM on $S^2$ expanded around the monopole 
background is equivalent to
$U(N)$ CS theory on $S^3$ in the planar limit.
Here, we can identify the difference of the 
monopole charges $n_s-n_t=s-t$ with the momenta
$n \in Z$ along the $S^1$ fiber direction and $\Lambda$
is interpreted as the momentum cutoff.
Introducing $N$ and taking the limit $N \rightarrow \infty$
are needed to suppress the non-planar contribution \cite{Ishii:2008ib}.
As for the relation between the coupling constants of the two theories, we naively expect
that 
$-\frac{g_{YM}^2AN}{2\pi}=\frac{N}{k}=\mbox{fixed}$. However, 
the last relation in (\ref{limit 1}) implies that we need some renormalization.
In sections 5 and 6, we show that the above prescription including
the renormalization of the coupling constant is indeed valid.

Furthermore, by combining this equivalence with the 
relationship between 2d YM on $S^2$ and
the matrix model reviewed in section 3.4, 
we obtain the following statement:
if one takes in (\ref{matrix background})
\begin{align}
&-\frac{\Lambda}{2} \leq s \leq \frac{\Lambda}{2}, \nonumber\\
&2j_s+1=\Omega+s, \nonumber\\
&N_s=N
\label{background}
\end{align}
and takes the limit in which
\begin{align}
&\Omega\rightarrow\infty, \;\;\; \Lambda\rightarrow\infty, \;\;\; \Omega-\frac{\Lambda}{2}\rightarrow\infty, \;\;\;
N\rightarrow\infty, \nonumber\\
&\frac{g_m^2N}{\Omega}=\frac{N}{4\pi (k+N)}=\mbox{fixed},
\label{S^3 limit}
\end{align}
the theory around (\ref{matrix background}) of the matrix model (\ref{N=1^* matrix model}) is equivalent 
to $U(N)$ CS theory on $S^3$ in the planar limit. In sections 5 and 6, we confirm this statement.
The particular background (\ref{background}) and the limit 
(\ref{S^3 limit}) is the same as the ones adopted in realizing
${\cal N}=4$ SYM on $R\times S^3$ in PWMM \cite{Ishii:2008ib}. 
$\Omega$ and $\Lambda$ play 
the role of the ultraviolet cutoffs for the angular momenta on $S^2$ and
in the fiber direction, respectively.

The results in section 3.4 and the above prescriptions lead 
us to consider the following statistical model,
\begin{align}
{\cal Z}=\int \prod_{s =-\Lambda/2}^{\Lambda/2}\prod_{i=1}^{N}dy_{si}
     \prod_{-\Lambda/2\leq s \leq t\leq \Lambda/2}
     \prod_{i,j=1}^{N}
\{ (y_{si}-y_{tj})^2+(s-t)^2 \}
e^{-\frac{1}{g^2}\sum_{s=-\Lambda/2}^{\Lambda/2}\sum_{i=1}^{N}y_{si}^2}.
\label{z of mm}
\end{align}
This model is obtained from (\ref{2dYM mp})
by making an analytic continuation as 
$y_{si} \rightarrow iy_{si}$, $g^2 \rightarrow -g^2$
and putting $n_s=s$, $N_s=N$ ($s= -\Lambda/2,\cdots ,\Lambda/2$) .
It follows from the above arguments that 
this model reproduces the planar limit of CS theory on $S^3$
in the limit in which
\begin{align}
\Lambda\rightarrow\infty,\;\;\;N\rightarrow\infty
\label{limit 2}
\end{align}
with
\begin{align}
2\pi^2g^2N=\frac{2\pi i N}{k+N}=g_sN=\mbox{fixed}.
\label{limit 3}
\end{align}
In the next section, we will verify 
the equivalence of CS theory on $S^3$
with (\ref{z of mm}) which is equivalent to 
the reduced matrix model expanded around the particular background.

We can easily extend the above argument to CS theory on $S^3/Z_q$.
We consider the case in which $Z_q$ acts on the circle along the 
fiber direction in $S^3$.
If we use the notation in appendix A, 
the periodicity for $S^3/Z_q$ is expressed as,
\begin{align}
(\theta,\varphi,\psi)\sim (\theta,\varphi+2\pi,\psi+2\pi)
\sim (\theta,\varphi,\psi+4\pi / q),
\end{align}
so that the radius of the fiber direction is $1/q$ of 
that for $q=1$ and the momenta along the fiber direction are 
discretized as $0, \pm q, \pm 2q, \cdots$. 
Then, we see that if one imposes (\ref{background})
with the second equation replaced with
$2j_s+1=\Omega+qs$ and takes the limit (\ref{S^3 limit}),
the theory around (\ref{matrix background}) of the matrix model 
(\ref{N=1^* matrix model}) is equivalent 
to $U(N)$ CS theory on $S^3/Z_q$ in the planar limit.
Correspondingly, the statistical model for CS theory
on $S^3/Z_q$ 
is given by replacing $s,t$ in (\ref{z of mm})
with $qs, qt$.

\section{Proof of equivalence}
In this section, we prove our statement that the model 
(\ref{z of mm}) reproduces the planar limit of CS theory on 
$S^3$ in the limit shown in (\ref{limit 2}) and (\ref{limit 3}).
This equivalence can be understood from a correspondence 
between the Feynman diagrams in the two theories.
The Feynman diagrams in (\ref{z of mm}) have a
one-to-one correspondence to those in (\ref{CSM}) which 
was obtained by rewriting the partition function of CS 
theory on $S^3$. Furthermore, 
if one takes the limit given by
(\ref{limit 2}) and (\ref{limit 3}), 
each Feynman diagram in (\ref{z of mm}) takes the
same value as the corresponding diagram in (\ref{CSM}).
We consider the free energy and the vacuum expectation value 
of unknot Wilson loop operator in the model (\ref{z of mm})
to see the equivalence.
We show that they coincide with the results 
in CS theory on $S^3$ completely in the planar limit.

\subsection{Feynman rule for our matrix model}
\label{Feynman rule for our matrix model}
Let us first consider the Feynman rule of (\ref{z of mm}).
For this purpose,
we rewrite (\ref{z of mm}) into a manifestly $U(N)$ invariant form.
We will see that the result is given by a multi-matrix model with 
double trace interaction.
This form allows us to compare the two theories explicitly.

We first factorize the measure in (\ref{z of mm}) as follows,
\begin{align}
\prod_{s \leq t} \prod_{i,j} \{(y_{si}-y_{tj})^2+(s-t)^2\}
=\sum_{s<t} (s-t)^{2N^2} \prod_s \prod_{i<j} (y_{si}-y_{sj})^2
\prod_{s<t}\prod_{i,j} \left( 
1+ \frac{(y_{si}-y_{tj})^2}{(s-t)^2}\right).
\label{ms in mm}
\end{align}
The first factor on the right hand side is just a constant 
which we will omit in the following and
the second factor gives the Vandermonde determinant for each $s$.
We next introduce the 't Hooft coupling in this model as,
\begin{align}
\lambda= 2\pi^2g^2N,
\label{tHooft in MM}
\end{align}
and redefine the fields as 
$\phi_{si}\equiv 2\pi y_{si}/\sqrt{\lambda}$.
Then, the third factor in (\ref{ms in mm}) can be written as
\begin{align}
&\prod_{s<t}\prod_{i,j} \left( 
1+ \frac{(y_{si}-y_{tj})^2}{(s-t)^2}\right)
\nonumber\\
&=\exp \left\{
\frac{1}{2}\sum_{s\neq t}\sum_{i,j}
\log \left( 
1+ \frac{\lambda}{4\pi^2}\frac{(\phi_{si}-\phi_{tj}^2)}{(s-t)^2}
\right)
\right\}
\nonumber\\
&=\exp \left\{
-\sum_{s\neq t}\sum_{i,j} \sum_{k=1}^{\infty}
\frac{1}{2k(s-t)^{2k}}
\left(\frac{-\lambda}{4\pi^2}\right)^k 
\sum_{m=0}^{2k}
\left( 
\begin{array}{c}
2k \\
m  \\
\end{array}
\right)  \phi_{si}^m
(- \phi_{tj})^{2k-m}
\right\}.
\label{ms in mm 2}
\end{align}
Finally, defining $U(N)$ covariant matrices $\phi_{s}$ such that
$\phi_{si}$ is $i$-th eigenvalue of $\phi_s$ for each $s$,
we find that (\ref{z of mm}) can be written as a multi-matrix model,
\begin{align}
{\cal Z}=\int \prod_{s}d\phi_s \exp \left\{
-\frac{N}{2} \sum_{s} {\rm tr} \phi_s^2 -\tilde{V}(\phi_s)
\right\},
\label{mm covariant}
\end{align}
where $\tilde{V}(\phi_s)$ is given by the double trace interaction,
\begin{align}
\tilde{V}(\phi_s)=\sum_{s\neq t}\sum_{k=1}^{\infty}
\frac{1}{2k(s-t)^{2k}}
\left(\frac{-\lambda}{4\pi^2}\right)^k 
\sum_{m=0}^{2k}
\left( 
\begin{array}{c}
2k \\
m  \\
\end{array}
\right)
{\rm tr} \phi_s^m
{\rm tr}(- \phi_t)^{2k-m}.
\label{V in mm}
\end{align}

\begin{figure}[tbp]
\begin{center}
\includegraphics[height=2.5cm, keepaspectratio, clip]{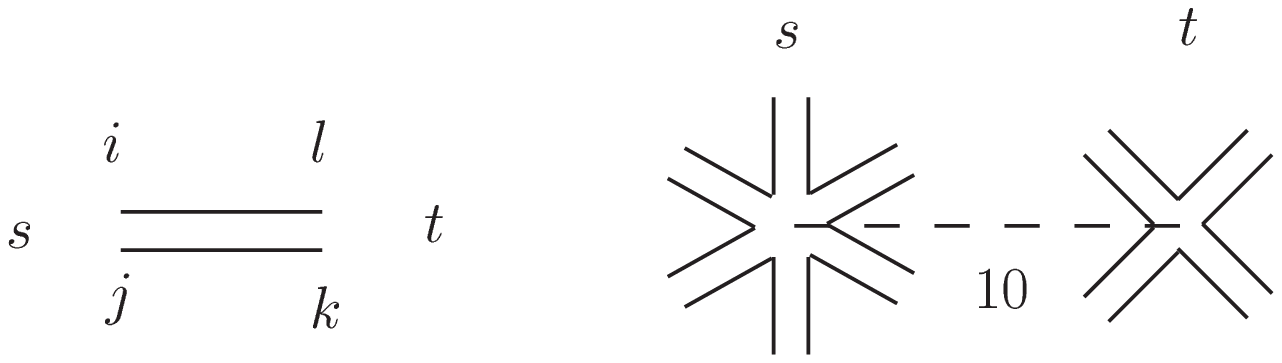}
\end{center}
\caption{The propagator (left) and the interaction
vertex (right) in the model (\ref{z of mm}). 
The right figure shows
the vertex which includes ${\rm tr}\phi_s^6 {\rm tr}\phi_t^4$.
The number $10$ represents the total number of $\phi_s$ and $\phi_t$
included in the vertex.}
\label{fig vertices in mm}
\end{figure}
The Feynman rule for (\ref{mm covariant}) is 
given as follows.
The tree level propagator is 
\begin{align}
\langle \phi_{sij} \phi_{tkl} \rangle 
= \frac{1}{N}\delta_{il}\delta_{jk}\delta_{st}.
\end{align}
In terms of the standard double line notation,
this is represented by double lines with indeces $s$ and $t$
as shown in Fig.\ref{fig vertices in mm}.
The double trace interaction vertices are represented in terms of
dashed lines which connect two traces.
For example, Fig.\ref{fig vertices in mm} (right) shows the term in 
(\ref{V in mm}) which includes ${\rm tr}\phi_s^6 {\rm tr}\phi_t^4$.
We connect two vertices coming from ${\rm tr}\phi_s^6$ and ${\rm tr}\phi_t^4$
in terms of a dashed line to represent a double trace interaction vertex.
The weights of the vertices can be easily read off from 
(\ref{V in mm}).
The ${\rm tr}\phi_s^a {\rm tr}\phi_t^b$ vertex gives a constant factor,
\begin{align}
-\frac{4}{a+b} \frac{1}{2(s-t)^{a+b}}
\left(\frac{-\lambda}{4\pi^2} \right)^{\frac{a+b}{2}}
(-1)^a
\left( 
\begin{array}{c}
a+b \\
a  \\
\end{array}
\right).
\label{vertex in mm}
\end{align}

\begin{figure}[tbp]
\begin{center}
\includegraphics[height=2.6cm, keepaspectratio, clip]{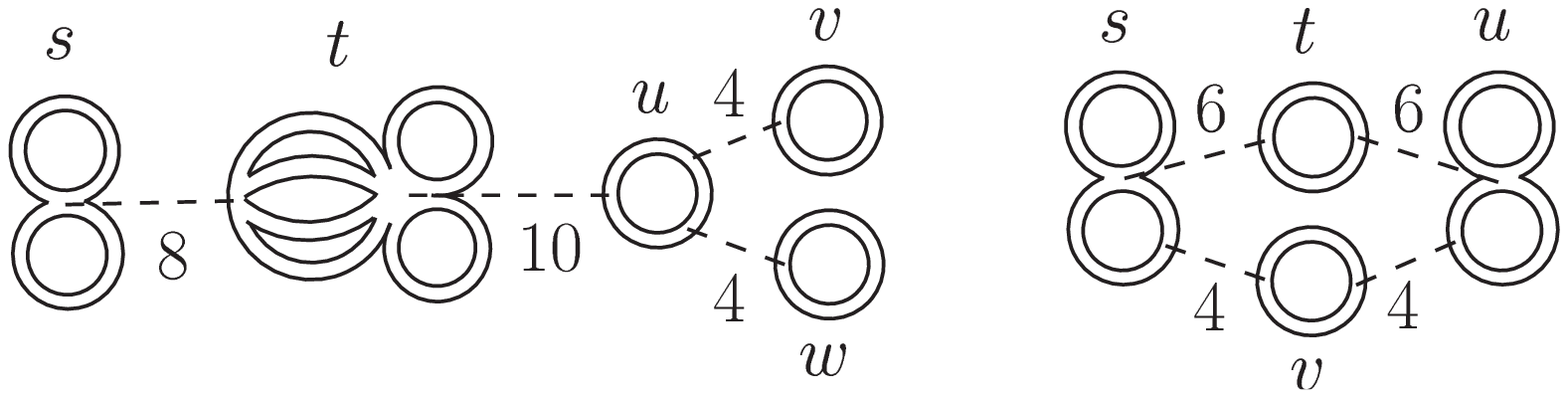}
\end{center}
\caption{Typical connected Feynman diagrams which contribute to 
the free energy in the model (\ref{mm covariant}).}
\label{typical diagram in mm}
\end{figure}
As in the standard perturbation theory, physical observables 
such as the free energy are computed by summing all the connected
diagrams.
Fig.\ref{typical diagram in mm} 
shows typical examples of connected diagrams 
which contribute to the free energy.
More precisely, the connected diagrams in this theory 
are those which are connected by dashed lines or by double lines.
One can show that only such diagrams indeed contribute to the
calculation of observables.

Since we are interested in the planar limit, 
we consider what type of diagrams is dominant in this limit.
It is easy to see that the leading order in the 
$1/N$ expansion
is given by the diagrams that satisfy two 
conditions.
One is that they are planar diagrams 
with respect to the double lines in the ordinary sense.
The other is that they are divided by two
parts by cutting any dashed line.
We call the diagrams satisfying the latter condition
`tree' diagrams in this paper.
We can see that the diagrams in which the dashed lines 
form any loop (i.e. non-`tree' diagrams) give subleading contribution.
In Fig.\ref{typical diagram in mm},
the left diagram contributes to the free energy
in the large-$N$ limit while
the right diagram is suppressed
since the dashed lines form a loop.

\subsection{Feynman rule for Chern-Simons matrix model}
Next, we construct a Feynman rule of CS matrix model (\ref{CSM}).
We first rewrite (\ref{CSM}) into a manifestly $U(N)$
invariant form as we did for (\ref{z of mm}) in the previous subsection.
The identification for the coupling constants shown in 
(\ref{limit 3}) allows us to put
\begin{align}
\lambda = g_sN,
\label{tHooft in CSM}
\end{align}
where $\lambda$ is the 't Hooft coupling in (\ref{mm covariant})
which is introduced in (\ref{tHooft in MM}).
If we redefine the fields as $\phi_i \equiv \beta_i/\sqrt{\lambda}$.
we can rewrite the measure in (\ref{CSM}) as follows,
\begin{align}
\prod_{i<j} \left( 2\sinh \frac{\beta_i-\beta_j}{2} \right)
&\sim \prod_{i<j}(\phi_i-\phi_j)^2 
\exp\left\{
{\sum_{i, j} \log 
\frac{\sinh \frac{\sqrt{\lambda}}{2}(\phi_i-\phi_j)}
{\frac{\sqrt{\lambda}}{2}(\phi_i-\phi_j)}}
\right\} 
\nonumber\\
&=
\prod_{i<j}(\phi_i-\phi_j)^2 
\exp \left\{
\sum_{i, j} \sum_{n=1}^{\infty} \log 
\left( 1+ \frac{\lambda}{4\pi^2n^2}(\phi_i-\phi_j)^2 \right)
\right\}
\nonumber\\
&=
\prod_{i<j}(\phi_i-\phi_j)^2 
\exp\left\{
\sum_{i, j} \sum_{k=1}^{\infty} \frac{-\zeta(2k)}{k}
\left(\frac{-\lambda}{4\pi^2} \right)^k
\sum_{m=0}^{2k}  
\left( 
\begin{array}{c}
2k \\
m  \\
\end{array}
\right)
\phi_i^m(-\phi_j)^{2k-m}
\right\},
\label{ms}
\end{align}
where $\sim$ represents that we have omitted an irrelevant 
constant factor and $\zeta(z)$ is the zeta function. 
We have used the formula,
\begin{align}
\frac{\sinh \pi x}{\pi x}= \prod_{n=1}^{\infty} 
\left( 1+ \frac{x^2}{n^2} \right),
\label{sinh formula}
\end{align}
to obtain the second line in (\ref{ms}).
If we introduce an $U(N)$ covariant matrix $\phi$ such that
its $i$-th eigenvalue is given by $\phi_i$, 
we obtain the following partition function 
for CS matrix model on $S^3$ \cite{Aganagic:2002wv}, 
\begin{align}
{\cal Z}_{CSM} \sim \int d\phi 
\exp \left\{ -\frac{N}{2}{\rm tr}\phi^2 -V(\phi) \right\},
\label{CSM covariant}
\end{align}
where the potential $V(\phi)$ is given by the 
double trace interaction,
\begin{align}
V(\phi)=\sum_{k=1}^{\infty}
\frac{\zeta(2k)}{k}
\left(\frac{-\lambda}{4\pi^2} \right)^k
\sum_{m=0}^{2k}  
\left( 
\begin{array}{c}
2k \\
m  \\
\end{array}
\right)
{\rm tr}\phi^m
{\rm tr}(-\phi)^{2k-m}.
\label{V in CSM}
\end{align}

We construct the Feynman rule of (\ref{CSM covariant}) as follows.
The tree level propagator is given by 
\begin{align}
\langle \phi_{ij} \phi_{kl} \rangle = \frac{1}{N}\delta_{il}\delta_{jk},
\end{align}
which is expressed by double lines as usual in the standard 
double line notation.
The double trace interaction vertices are represented in terms of 
dashed lines which connect two traces as in the case of 
(\ref{mm covariant}).
They are represented in the same way as in Fig.\ref{fig vertices in mm}
except that we do not need the indeces $s,t$ here.
Fig.\ref{fig vertices in mm} (right) without $s,t$
corresponds to the term in 
(\ref{V in CSM}) which includes ${\rm tr}\phi^6 {\rm tr}\phi^4$.
The weights of the vertices can be easily read off from 
(\ref{V in CSM}).
The ${\rm tr}\phi^a {\rm tr}\phi^b$ vertex gives a constant factor,
\begin{align}
-\frac{4 \zeta(a+b)}{a+b}
\left(\frac{-\lambda}{4\pi^2} \right)^{\frac{a+b}{2}}
(-1)^a
\left( 
\begin{array}{c}
a+b \\
a  \\
\end{array}
\right).
\label{vertex in csmm}
\end{align}

As in the case of (\ref{mm covariant}), connected diagrams
in this theory are those which are connected by dashed lines or
by double lines. 
In the $N\rightarrow \infty$ limit, 
the leading order in the $1/N$ expansion is given by 
the diagrams which are planar in the ordinary sense and 
`tree' with respect to dashed lines.
For instance, the right 
diagram in Fig.\ref{typical diagram in mm} 
in the case of (\ref{CSM covariant}) 
(i.e. the diagram without the indeces $s,t,u,\cdots$.)
is suppressed in the large-$N$ limit since
the dashed lines form a loop.
Note that the Feynman diagrams in (\ref{CSM covariant})
have the one-to-one 
correspondence to those in (\ref{mm covariant}).


\subsection{Equivalence between two theories}
In this subsection, 
we show that (\ref{mm covariant}) and (\ref{CSM covariant}) are equivalent
in the limit shown in (\ref{limit 2}) and (\ref{limit 3}).
We can prove this equivalence by showing that, in this limit,
each Feynman diagram in (\ref{mm covariant}) takes the same value
as the corresponding diagram in (\ref{CSM covariant}).
Note that we have already imposed the identification
of the coupling constants shown in (\ref{limit 3})
through (\ref{tHooft in MM}) and (\ref{tHooft in CSM}).

\subsubsection{Free energy}
In order to prove the equivalence,
we first show that
\begin{align}
\frac{\cal F}{N^2(\Lambda+1)} =
\frac{{\cal F}_M}{N^2}
\label{free energy agreement}
\end{align}
holds in the limit shown in (\ref{limit 2})
and (\ref{limit 3}),
where ${\cal F}$ and ${\cal F}_M$ are
the free energies of 
(\ref{mm covariant}) and (\ref{CSM covariant}),
respectively.
As an example of the Feynman diagrams 
contributing to the free energy, 
let us consider the left diagram in 
Fig.\ref{typical diagram in mm} for the both theories.
Since the only deference between (\ref{vertex in mm}) and 
(\ref{vertex in csmm}) is  
$\frac{1}{2(s-t)^{a+b}}$ and $\zeta(a+b)$, 
it is easy to see that the diagram in (\ref{mm covariant}) 
divided by $N^2(\Lambda+1)$
becomes equal to the corresponding diagram in (\ref{CSM covariant}) 
divided by $N^2$ if
\begin{align}
&\frac{1}{2^4(\Lambda+1)}
\sum_{s=-\Lambda /2}^{\Lambda /2}
\sum_{ \substack{t=-\Lambda /2 \\ (t\neq s)} }^{\Lambda /2}
\sum_{ \substack{u=-\Lambda /2 \\ (u\neq t)} }^{\Lambda /2}
\sum_{ \substack{v=-\Lambda /2 \\ (v\neq u)} }^{\Lambda /2}
\sum_{ \substack{w=-\Lambda /2 \\ (w\neq u)} }^{\Lambda /2}
\frac{1}{(s-t)^8(t-u)^{10}(u-v)^4(u-w)^4}
\nonumber\\
&=\zeta(8)\zeta(10)\zeta(4)^2
\label{zeta identity}
\end{align}
holds.
In fact, this identity holds in the
$\Lambda \rightarrow \infty$ limit. We 
show the proof of (\ref{zeta identity}) in appendix
\ref{Proof of identity}. 
Thus, we see that the two diagrams with the appropriate 
normalization shown in (\ref{free energy agreement})
take the same value in the $\Lambda \rightarrow \infty$ limit.

\begin{figure}[tbp]
\begin{center}
\includegraphics[height=3.5cm, keepaspectratio, clip]{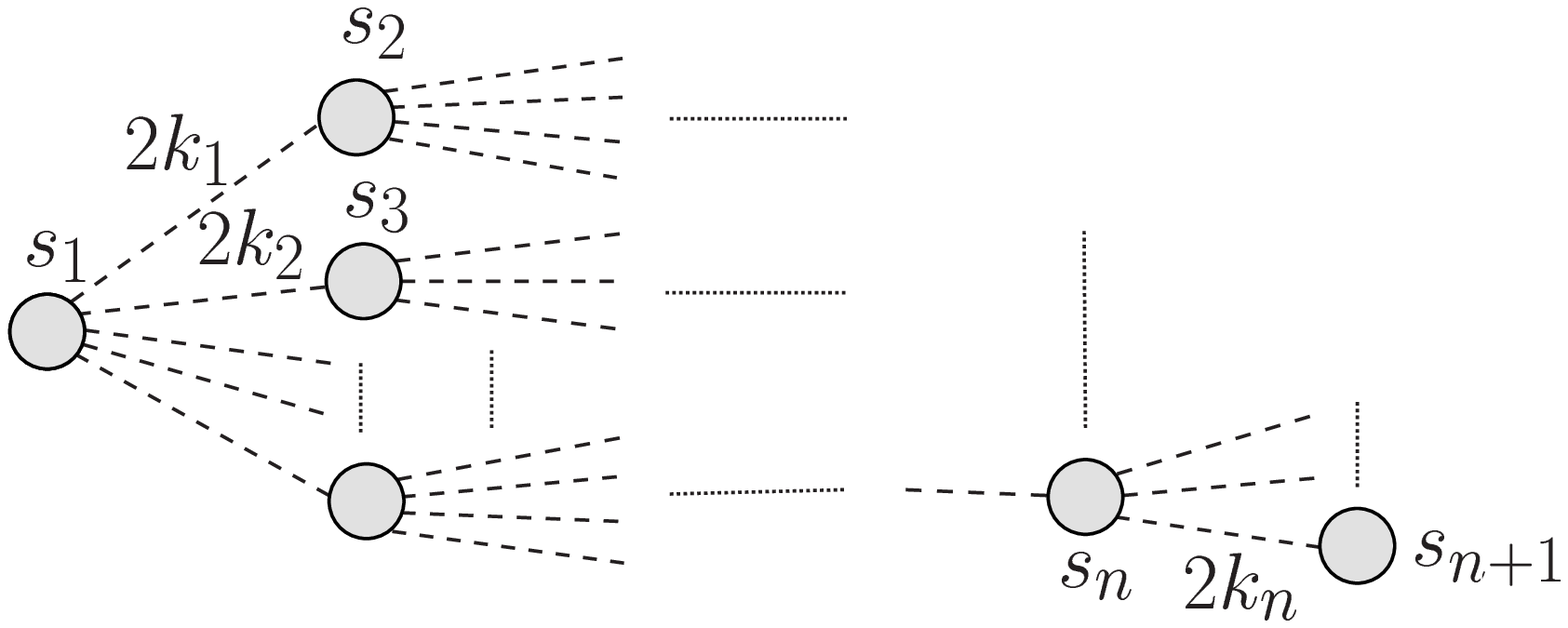}
\end{center}
\caption{Generic `tree' planar diagram in
(\ref{mm covariant}). Each blob represents
a planar diagram.}
\label{tree planar diagram}
\end{figure}

We can show more general formula which 
includes (\ref{zeta identity}) as a special case.
Let us consider generic 
`tree' planar diagram in (\ref{mm covariant})
shown in Fig.\ref{tree planar diagram}.
For this diagram, we can show the following 
identity in the large-$\Lambda$ limit\footnote{
Note that (\ref{zeta identity 2}) holds also for 
the diagrams which are `tree' and non-planar
since (\ref{zeta identity 2}) depends only on 
the structure of the dashed lines.},
\begin{align}
\frac{1}{2^n(\Lambda+1)}
\sum_{\{s_a\}}
\frac{1}{(s_1-s_2)^{2k_1}(s_1-s_3)^{2k_2}\cdots (s_n-s_{n+1})^{2k_n}}
=\prod_{a=1}^{n}\zeta(2k_a),
\label{zeta identity 2}
\end{align}
where $\{s_a\}$ represents the set 
of indeces $(s_1, \cdots, s_{n+1})$ and each $s_a$
runs from $-\Lambda/2$ to $\Lambda/2$.
Note that the singular points such as $s_1=s_2$ are 
not included in the summations as in the simpler
case of (\ref{zeta identity}).
The proof of (\ref{zeta identity 2}) 
is shown in appendix \ref{Proof of identity}.
(\ref{zeta identity 2}) implies that
the value of the diagram in
Fig.\ref{tree planar diagram} divided by
$N^2(\Lambda+1)$ is equal to the 
same diagram in (\ref{CSM covariant}) divided
by $N^2$.
We find, therefore, that (\ref{free energy agreement}) holds to 
all orders in $\lambda$ in the limit (\ref{limit 2}).

We can check (\ref{free energy agreement})
also by performing an explicit perturbative 
calculation.
The detail of the calculation is shown in 
appendix \ref{Perturbative calculation}.
Up to ${\cal O}(\lambda^3)$,
the result is given by,
\begin{align}
\frac{\cal F}{N^2(\Lambda+1)}
=&
\frac{\lambda}{12}
+\frac{\lambda^2}{288}
+\frac{\lambda^2}{1440N^2}
\nonumber\\
&+\frac{\lambda^3}{N^2}
\left(
\frac{1}{9072}+\frac{1}{48\pi^6 \Lambda}
\sum_{s,t,u}
\frac{1}{(s-t)^2(t-u)^2(u-s)^2}
\right)
+{\cal O}(\lambda^4),
\label{pc 1}
\end{align}
while for CS matrix model,
we can read off the counterpart from (\ref{free energy in CS})
as,
\begin{align}
\frac{{\cal F}_M}{N^2}=
\frac{\lambda}{12}
-\frac{\lambda}{12N^2}
+\frac{\lambda^2}{288}
+\frac{\lambda^2}{96N^2}
+{\cal O}(\lambda^4).
\label{pc 2}
\end{align}
Thus, we see that if one takes 
the limit shown in (\ref{limit 2}) and (\ref{limit 3}),
(\ref{pc 1}) indeed agrees with (\ref{pc 2}).

\subsubsection{Unknot Wilson loop}
In terms of the formula (\ref{zeta identity 2}),
we can show that other physical observables also
take the same values in the two theories.
In the following, 
we show that the VEV of unknot Wilson loop 
in CS theory which is given by 
(\ref{WL in noncanonical framing})
is also reproduced from the model (\ref{z of mm}).
%
%

We introduce the Wilson loop operator in ${\cal N}=1^*$
matrix model proposed in \cite{Ishii:2007sy},
\begin{align}
\hat{W}({\cal K})=\frac{1}{M}{\rm Tr} \left[
P\exp \left( i\mu \int^1_0 X_iE^i_M(z(\sigma))
\frac{dz^M(\sigma)}{d\sigma}d\sigma
\right)
\right],
\label{WL in MM}
\end{align}
where $z^M$ are coordinates on $S^3$,
$\sigma$ parameterize the knot ${\cal K}$ and 
$E^i_Mdz^M$ are the right-invariant 1-forms on $S^3$ 
defined in (\ref{right invariant 1-form}). 
In \cite{Ishii:2007sy}, it is shown through Taylor's T-duality
that this operator is reduced to the Wilson loop
operator on $S^3$ in the continuum limit.
In order to see the correspondence to
the unknot Wilson loop in CS theory,
we consider (\ref{WL in MM}) with an unknot contour.
We take the simplest one, a great circle on $S^3$,
as such contour.
The great circle is parameterized, for example, by 
\begin{align}
(\theta, \varphi, \psi)=\left(\frac{\pi}{2},0, 4\pi\sigma\right),
\label{great circle}
\end{align}
with $\sigma \in (0,1]$.
By substituting (\ref{great circle}) into (\ref{WL in MM})
and setting $\mu=1$,
we obtain,
\begin{align}
\hat{W}({\rm unknot})=\frac{1}{M}{\rm Tr} \left(
e^{ 4\pi i X_3 }
\right).
\label{WL 0}
\end{align}
As in section 3.3, we take the gauge in which $X_3=\Phi$ is
diagonalized. 
Recall that after we integrate out the delta functions 
shown in (\ref{product of delta functions}) 
which appear in the partition function,
the remaining eigenvalues of $\Phi$ are represented by $y_{si}/2 $.
Then, in (\ref{WL 0}),
if we make the analytic continuation $y_{si}\rightarrow -iy_{si}$ and
the field redefinition $y_{si}=\sqrt{\lambda}\phi_{si}/{2\pi}$,
we obtain,
\begin{align}
\hat{W}({\rm unknot})=
\frac{\Omega}{M}\sum_{s,i}e^{\sqrt{\lambda} \phi_{si}}
=\frac{1}{N(\Lambda+1)}\sum_{s}{\rm Tr}e^{\sqrt{\lambda}\phi_s}.
\label{WL}
\end{align}
Therefore, we find that the unknot Wilson loop operator
in the fundamental representation 
is given by (\ref{WL}) in (\ref{z of mm}).

On the other hand, in CS matrix model (\ref{CSM}),
the VEV of unknot Wilson loop is given by
(\ref{Wilson loop in non-canonical framing}).
Then, the equivalence for the unknot Wilson loop is 
expressed in terms of $\phi$ and $\phi_s$ as
\begin{align}
\frac{1}{N(\Lambda+1)}\sum_{s} 
\langle {\rm Tr}e^{\sqrt{\lambda}\phi_s}
\rangle
=
\frac{1}{N} 
\langle 
{\rm Tr} e^{\sqrt{\lambda}\phi}
\rangle_M,
\label{equivalence for Wilson loop}
\end{align}
where $\langle \cdots \rangle$ and 
$\langle \cdots \rangle_M$ denote the expectation
values in (\ref{mm covariant}) and (\ref{CSM covariant}),
respectively.
We can show
(\ref{equivalence for Wilson loop})
in all orders in perturbation theory.
In perturbation theory, the VEVs in 
(\ref{equivalence for Wilson loop}) 
are calculated by expanding the exponentials.
Then, (\ref{equivalence for Wilson loop}) is satisfied
if 
\begin{align}
\frac{1}{\Lambda+1}
\sum_{s}
\langle 
{\rm Tr} \phi^n_s
\rangle
=\langle 
{\rm Tr} \phi^n
\rangle_M
\label{equivalence for Wilson loop 2}
\end{align}
holds for $n=0,1,2,\cdots$.
Because
(\ref{zeta identity 2}) holds for 
all the `tree' planar Feynman diagrams 
contributing to the VEVs in
(\ref{equivalence for Wilson loop 2}),
we find that 
(\ref{equivalence for Wilson loop 2}) and hence
(\ref{equivalence for Wilson loop}) hold 
in all orders in $\lambda$.
Thus, we have shown that the VEV of unknot 
Wilson loop in CS theory is 
reproduced from the reduced model.

It is straightforward to check
(\ref{equivalence for Wilson loop}) explicitly by
calculating the VEV of (\ref{WL}) 
in perturbation theory.
The method of calculating the 
VEV is almost the same as in the case of the free energy.
We show the detail of this calculation in appendix
\ref{Perturbative calculation}.
In the large-$\Lambda $ limit,
the result is given by
\begin{align}
\frac{1}{N(\Lambda+1)}\sum_{s} 
\langle {\rm Tr}e^{\sqrt{\lambda}\phi_s}
\rangle
=1+\frac{\lambda}{2}+\frac{\lambda^2}{6}
  +\frac{\lambda^2}{24N^2}
  +\frac{\lambda^3}{24}
  +\frac{7\lambda^3}{240N^2}
  +{\cal O}(\lambda^4).
\label{VEV of Wilson loop in matrix model}
\end{align}
On the other hand, (\ref{WL in noncanonical framing})
is expanded as,
\begin{align}
\langle W_{\square}({\rm unknot}) \rangle_{CS}
&=
1+\frac{\lambda}{2}
+\frac{\lambda^2}{6}
-\frac{\lambda^2}{24N^2}
+\frac{\lambda^3}{24}
-\frac{\lambda^3}{48N^2} +{\cal O}(\lambda^4).
\label{unknot wilson loop in cs}
\end{align}
Comparing (\ref{VEV of Wilson loop in matrix model})
and (\ref{unknot wilson loop in cs}), 
we can see that
(\ref{equivalence for Wilson loop}) indeed holds
in the $N\rightarrow \infty$ limit.

\subsection{Extension to $S^3/Z_q$}
We can show the similar equivalence for the 
case of CS theory on $S^3/Z_k$.
Although this theory possesses many nontrivial vacua,
we consider the trivial vacuum only
in this paper.

The vacua in CS theory on $S^3/Z_q$ are
characterized by the holonomy along the circle on which $Z_q$ acts.
In \cite{Aganagic:2002wv}, it is shown that the 
partition function of CS theory on $S^3/Z_q$ 
for each vacuum sector is 
reduced to the partition function of a matrix model
similar to (\ref{CSM covariant}).
For example,
the partition function for the trivial vacuum 
sector is given by
(\ref{CSM covariant}) with $\lambda$ replaced 
by $\lambda'\equiv g_s N/q$.

On the other hand, as mentioned in the last part in
section \ref{CS theory on $S^3$ from the matrix model},
the statistical model for CS theory on $S^3/Z_q$
is given by replacing $s,t$ in (\ref{z of mm})
with $qs, qt$. By following the same calculation 
as (\ref{ms in mm}) and (\ref{ms in mm 2}), 
we can see that such model is equivalent to
(\ref{mm covariant}) with $\lambda$ replaced by
$\lambda '' \equiv 2\pi^2 g^2 N/q^2$.

Then, with the identification $\lambda'=\lambda''$,
we see that the two theories are equivalent in 
the limit (\ref{limit 2}).
Thus, we find that the theory on $S^3/Z_q$ is 
also reproduced from the matrix model.

\section{Monte Carlo simulation}

The large-$N$ equivalence between (\ref{z of mm})
and CS theory on $S^3$ can be 
understood also from the agreement of the 
eigenvalue densities in the two theories.
In this section, 
by performing a Monte Carlo simulation, we show that
the eigenvalue density of $y_{si}$ at the saddle point 
of (\ref{z of mm})
coincides with that of $\beta_i$ in (\ref{CSM}) if $|s|$ is
sufficiently small compared to the cutoff $\Lambda/2$.
The eigenvalue densities for $y_{si}$ with $s$ near the 
cutoff have some deviation from the density of $\beta_i$.
However, we will see that such cutoff effect vanishes 
in the continuum limit and does not contribute
to values of physical observables.
From this property of the eigenvalue density,
we can also give another check of 
the equivalences for free energy
and unknot Wilson loop operator which we have shown in the
previous section.

\begin{figure}[tbp]
\begin{center}
\rotatebox{-90}{
\includegraphics[height=11cm, keepaspectratio, clip]{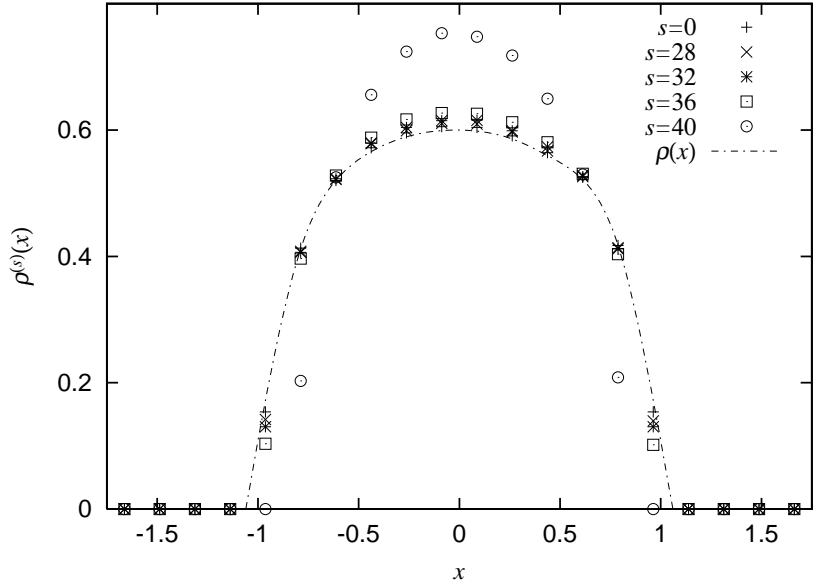}}
\end{center}
\caption{The eigenvalue distribution of $y_{si}$ in our model
is plotted for $s=0,28,32,36,40$ with $\lambda=0.25$, $\Lambda =80$ and $N=48$.
The dashed line represents the distribution $\rho(x)$
in CS theory on $S^3$. The error bars are omitted since 
they are smaller than the symbol size}
\label{fig.1}
\end{figure}

Let us introduce the eigenvalue density of $\beta_i$
in CS theory on $S^3$ as,
\begin{align}
\rho(x)=\frac{1}{N}\sum_{i=1}^N
\delta (x-\beta_{i}/(2\pi )),
\label{eigenvalue density in cs}
\end{align}
and the eigenvalue density of $y_{si}$ in (\ref{z of mm}) as,
\begin{align}
\rho^{(s)}(x)=\frac{1}{N}\sum_{i=1}^N
\delta (x-y_{si}).
\label{eigenvalue density in mm}
\end{align}
We also define the difference between (\ref{eigenvalue density in cs}) 
and (\ref{eigenvalue density in mm}) as,
\begin{align}
 \Delta \rho^{(s)}(x)=\rho^{(s)} (x) - \rho(x).
\label{delta rho}
\end{align}
By performing a Monte Carlo simulation of (\ref{z of mm}),
we can see that $\Delta \rho^{(s)}(x)=0$ for $s\sim {\cal O}(1)$ 
while $\Delta \rho^{(s)}(x)\neq 0$ if $s$ is near the cutoff $\pm \Lambda/2$.
Figure \ref{fig.1} shows the result of the simulation.
This result shows that as one goes toward the midpoint $s=0$,
the distribution converges rapidly to the distribution of 
CS theory on $S^3$. 
From this result, we conclude that the deviation of the eigenvalue density
exists only for sufficiently large $|s|$.

\begin{figure}[tbp]
\begin{center}
\hspace{-2cm}
\rotatebox{-90}{
\includegraphics[height=10cm, keepaspectratio, clip]{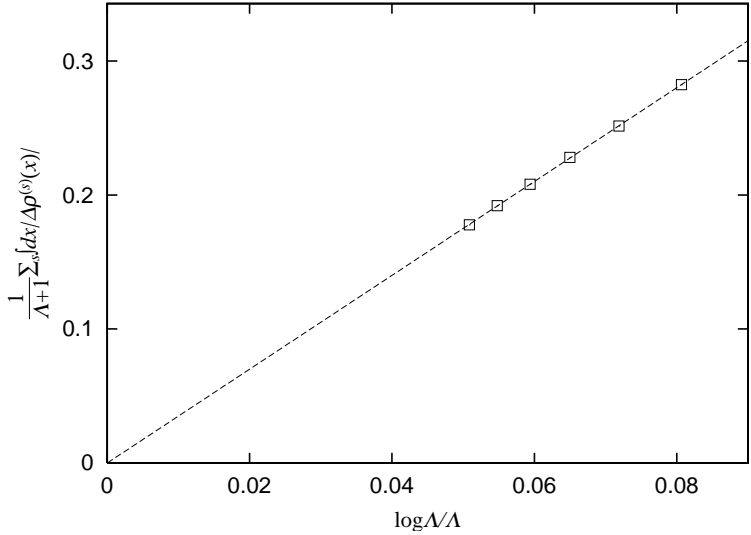}}
\end{center}
\caption{$\frac{1}{\Lambda +1}\int dx \sum_{s=-\Lambda/2}^{\Lambda/2}
|\Delta \rho^{(s)}(x)|$ is plotted against $\log \Lambda/\Lambda$.
$\lambda=0.25$ and $N$ has already been extrapolated 
from $N=24,32,48$ to $N=\infty$ by a linear function of $1/N^2$.
The dotted line represents the extrapolation of the plots
using $f(x)=a+b\frac{\log \Lambda}{\Lambda}$ with $a$ and $b$ constants.
The error bars are smaller than the symbol size.
}
\label{fig.2}
\end{figure}

We have also verified through the Monte Carlo simulation
that $\Delta \rho^{(s)}(x)$ possesses the following property,
\begin{align}
\lim_{\Lambda \rightarrow \infty}
\frac{1}{\Lambda+1}\int dx \sum_{s=-\Lambda/2}^{\Lambda/2}
|\Delta \rho^{(s)}(x)| =0.
\label{limit of delta rho}
\end{align}
This fact is shown as follows. 
Let us consider the saddle point equation of (\ref{z of mm}),
\begin{align}
\frac{2}{g^2}x-\int dx' \sum_{t=-\Lambda/2}^{\Lambda/2} \rho^{(t)}(x')
\left\{
\frac{1}{x-x'+i(s-t)}+\frac{1}{x-x'-i(s-t)}
\right\}=0.
\label{saddle pt eq}
\end{align}
Because 
$\Delta \rho^{(s)}(x)=0$ for $s\sim {\cal O}(1)$, 
in order for (\ref{saddle pt eq}) to be satisfied,
$\Delta \rho^{(s)}(x)$ with $s$ near the cutoff
should behave as
\begin{align}
\Delta \rho^{(s)}(x) \sim \frac{1}{\Lambda/2-|s|}.
\label{behavior of delta rho}
\end{align}
Then, we can expect that
\begin{align}
\frac{1}{\Lambda+1}\int dx \sum_{s=-\Lambda/2}^{\Lambda/2}
|\Delta \rho^{(s)}(x)| \sim \frac{\log \Lambda}{\Lambda},
\label{logT/T}
\end{align}
in the limit $\Lambda \rightarrow \infty$ so that
(\ref{limit of delta rho}) is satisfied.
We confirmed (\ref{logT/T}) through the
Monte Carlo simulation.
Figure \ref{fig.2} is the result of the simulation
and it shows that 
the quantity on the left hand side in (\ref{logT/T})
is indeed nicely fitted by
$f(x)=a+b\frac{\log \Lambda}{\Lambda}$ with $a\simeq 0$.

If we assume (\ref{limit of delta rho}) or (\ref{logT/T}), 
we check that the equivalence of the free energy 
(\ref{free energy agreement}) indeed holds.
Free energy of (\ref{z of mm}) is written in terms of
(\ref{eigenvalue density in mm}) as,
\begin{align}
\frac{{\cal F}}{\Lambda N^2}
=-\frac{1}{g^2\Lambda}\sum_{s}\int dx x^2 \rho^{(s)}(x)
+\frac{1}{2\Lambda}\sum_{s\neq t} \int dx dx'
\rho^{(s)}(x) \log 
\{ (x-x')^2 + (s-t)^2 \}\rho^{(t)}(x').
\end{align}
Then, by substituting (\ref{delta rho}), we obtain
\begin{align}
\frac{{\cal F}}{\Lambda N^2}
&=-\frac{1}{g^2}\int dx x^2 \rho(x)
+\frac{1}{2\Lambda }\sum_{s\neq t} \int dx dx'
\rho (x) \log \{ (x-x')^2 + (s-t)^2 \}\rho (x')
\nonumber\\
&\;\; 
-\frac{1}{g^2\Lambda }\sum_{s}\int dx x^2 
\Delta \rho^{(s)}(x)
+\frac{1}{\Lambda }\sum_{s\neq t} \int dx dx'
\rho (x) \log \{ (x-x')^2 + (s-t)^2 \}
\Delta \rho^{(t)} (x')
\nonumber\\
&\;\;
+\frac{1}{2\Lambda }\sum_{s\neq t} \int dx dx'
\Delta \rho^{(s)} (x) 
\log \{ (x-x')^2 + (s-t)^2 \}
\Delta \rho^{(t)} (x').
\label{F and FCS}
\end{align}
By using the formula (\ref{sinh formula}),
we can show that 
the first line on the right hand side of (\ref{F and FCS}) 
is equal to ${\cal F}_{M}/N^2$ in CS matrix model
in the limit $\Lambda \rightarrow \infty$.
We can further show that the second and the third lines vanish
in $\Lambda \rightarrow \infty$ as follows.
Let us introduce a sufficiently large but finite 
constant $b$ such that the eigenvalue
densities for all $s$ fit in the interval 
$[-b,b]$.
Such $b$ exists because 
the attractive force in (\ref{saddle pt eq}) 
coming from the Gaussian potential
is stronger than the repulsive force for $|x| \gg 1$
so that $\rho^{(s)}(x)$ has a compact support.
Then, the first term in the second line is bounded as
\begin{align}
-\frac{1}{g^2\Lambda}\sum_{s=-\Lambda/2}^{\Lambda/2}\int dx x^2 
\Delta \rho^{(s)}(x)
\leq \frac{b^2}{g^2\Lambda }\int dx \sum_{s=-\Lambda/2}^{\Lambda/2} 
|\Delta \rho^{(s)}(x)|,
\end{align}
Because of (\ref{limit of delta rho}), we find that this term
does not contribute to ${\cal F}/(\Lambda N^2)$ in the 
$\Lambda \rightarrow \infty$ limit.
We can apply the same argument to all the terms in the
second and the third lines in (\ref{F and FCS}).
Thus, we have given another check of (\ref{free energy agreement}).

We can also show the equivalence for unknot Wilson loop operator.
Since the VEV in (\ref{z of mm}) can be written as
\begin{align}
\langle \hat{W}({\rm unknot}) \rangle
=\frac{\Omega}{M}\langle \sum_{s,i} e^{2\pi y_{si}} \rangle
=\frac{1}{N(\Lambda+1)}\int dx \sum_{s} 
\langle \rho^{(s)}(x) \rangle e^{2\pi x},
\end{align}
Using (\ref{limit of delta rho}), we find that
\begin{align}
\langle \hat{W} ({\rm unknot}) \rangle 
\rightarrow \frac{1}{N} \int dx \langle \rho(x) \rangle
e^{2\pi x} = \langle W_{\square}({\rm unknot}) \rangle_{CS},
\end{align}
in the limit $\Lambda \rightarrow \infty$.
Thus, we have given another derivation of the equivalence for
unknot Wilson loop.

\section{Conclusion and discussion}
In this paper, we studied a matrix model that is obtained 
by the dimensional reduction of CS theory on $S^3$ 
to zero dimension.
We found that expanded around the particular background 
corresponding to $S^3$, 
it reproduces the original theory on $S^3$ in the planar limit.
We calculated the partition function and the VEV of the 
Wilson loop in the theory around the background
of the matrix model and verified that they agree with those 
in the planar CS theory on $S^3$.
We checked these results also by performing the Monte Carlo
simulation.
We also extended this equivalence to the case of 
CS theory on $S^3/Z_q$.

In this paper, we only consider the unknot Wilson loop.
It is relevant to see whether the VEV of the Wilson loop for a 
real knot is reproduced in our formulation.
We should verify that the VEV of the Wilson loops does not change 
against continuous deformation of the loops, such
that it indeed represents a topological invariant.
We should also extend our analysis to the case of 
the Wilson loop with general representation
${\cal R}$ of $SU(N)$.
It is also interesting to see whether we can
use our formulation to achieve the large-$N$ reduction for 
other theories
including the CS term such as the ABJM model\cite{Hanada:2009hd}.


\section*{Acknowledgements}
We would like to thank K. Ohta for his collaboration in the early 
stage of this work and for many discussions.
The work of G.\ I.\ is supported by the National Research Foundation
of Korea(KRF) grant funded by the Korea government(MEST)
through the Center for Quantum Spacetime(CQUeST) of Sogang University
with grant number 2005-0049409.
The work of S.\ S.\ 
is supported in part by the JSPS Research Fellowship for Young Scientists. 
The work of A.\ T.\ is supported in part 
by Grant-in-Aid for Scientific Research (19540294) from JSPS.

\appendix
\section{$S^3$ and $S^2$}
\label{$S^3$ and $S^2$}

In this appendix, we summarize some useful facts about $S^3$ and $S^2$ 
(See also \cite{Ishii:2008tm,Ishii:2008ib}).
$S^3$ is viewed as the $SU(2)$ group manifold. We parameterize an element
of $SU(2)$ in terms of the Euler angles as
\begin{equation}
g=e^{-i\varphi \sigma_3/2}e^{-i\theta \sigma_2/2}e^{-i\psi \sigma_3/2},
\label{Euler angles}
\end{equation}
where $0\leq \theta\leq \pi$, $0\leq \varphi < 2\pi$, $0\leq \psi < 4\pi$.
The periodicity with respect to these angle variables is expressed as
\begin{align}
(\theta,\varphi,\psi)\sim (\theta,\varphi+2\pi,\psi+2\pi)\sim (\theta,\varphi,\psi+4\pi).
\label{periodicity on S^3}
\end{align}
The isometry of $S^3$ is $SO(4)=SU(2)\times SU(2)$, and these two
$SU(2)$'s act on $g$ from left and right, respectively. 
We construct the right-invariant 1-forms, 
\begin{equation}
dgg^{-1}=-i\mu E^i \sigma_i/2,
\label{right-invariant 1-form}
\end{equation}
where $E^i$ are explicitly given by
\begin{eqnarray}
&&E^1=\frac{1}{\mu}(-\sin \varphi d\theta + \sin\theta\cos\varphi d\psi),\nonumber\\
&&E^2=\frac{1}{\mu}(\cos \varphi d\theta + \sin\theta\sin\varphi
 d\psi),\nonumber\\
&&E^3=\frac{1}{\mu}(d\varphi + \cos\theta d\psi).
\label{right invariant 1-form}
\end{eqnarray}
The radius of $S^3$ is $2/\mu$. 
$E_i$ satisfy the Maurer-Cartan equation
\begin{equation}
dE^i-\frac{\mu}{2}\epsilon_{ijk}E^j\wedge E^k=0.\label{Maurer-Cartan}
\end{equation}
The metric is constructed from $E^i$ as
\begin{equation}
ds^2=E^iE^i=\frac{1}{\mu^2}\left(
d\theta^2+\sin^2\theta d\varphi^2 +(d\psi+\cos\theta d\varphi)^2\right).
\label{metric of S^3}
\end{equation}
The Killing vector dual to $E^i$ is given by
\begin{equation}
{\cal{L}}_i=-\frac{i}{\mu}E^M_i\partial_M,
\label{definition of Killing vector}
\end{equation}
where $M=\theta,\varphi,\psi$ and $E^M_i$ are inverse of $E^i_M$. 
The Killing vector is explicitly expressed as
\begin{eqnarray}
&&{\cal{L}}_1
=-i\left(-\sin\varphi\partial_{\theta}-\cot\theta\cos\varphi\partial_{\varphi}
+\frac{\cos\varphi}{\sin\theta}\partial_{\psi}\right),\nonumber\\
&&{\cal{L}}_2
=-i\left(\cos\varphi\partial_{\theta}-\cot\theta\sin\varphi\partial_{\varphi}
+\frac{\sin\varphi}{\sin\theta}\partial_{\psi}\right),\nonumber\\
&&{\cal{L}}_3=-i\partial_{\varphi}.\label{Killing vector}
\end{eqnarray}
From the Maurer-Cartan equation (\ref{Maurer-Cartan}), one can show that the Killing vector satisfies 
the $SU(2)$ algebra $[{\cal{L}}_i,{\cal{L}}_j]=i\epsilon_{ijk}{\cal{L}}_k$.

Next, let us regard $S^3$ as a $U(1)$ bundle over $S^2=SU(2)/U(1)$.
$S^2$ is parametrized by $\theta$ and $\varphi$ and covered with two local patches:
the patch I defined by $0\leq\theta <\pi$ and the patch II defined by $0<\theta\leq\pi$.
In the following, the upper sign represents the patch I while the lower sign represents the patch II.
The element of $SU(2)$ in (\ref{Euler angles}) is decomposed as 
\begin{align}
g=L\cdot h
\end{align}
with
\begin{align}
&L=e^{-i\varphi \sigma_3/2}e^{-i\theta \sigma_2/2}e^{\pm i\varphi \sigma_3/2}, \nonumber\\ 
&h=e^{-i(\psi\pm\varphi) \sigma_3/2}. 
\end{align}
$L$ represents an element of $S^2$, while $h$
represents the fiber $U(1)$.
The fiber direction is parametrized by $y=\psi\pm\varphi$.
The zweibein of $S^2$ is given by the $i=1,2$ components of the left-invariant 1-form, 
$-iL^{-1}dL=\mu e^i\sigma_i/2$, which takes the form
\begin{align}
&e^1=\frac{1}{\mu}(\pm\sin\varphi d\theta+\sin\theta\cos\varphi d\varphi), \nonumber\\
&e^2=\frac{1}{\mu}(-\cos\varphi d\theta \pm \sin\theta\sin\varphi d\varphi).
\end{align}
This zweibein gives the standard metric of $S^2$ with the radius $1/\mu$:
\begin{align}
ds^2=e^ie^i=\frac{1}{\mu^2}(d\theta^2+\sin^2\theta d\varphi^2).
\label{metric of S^2}
\end{align}
Making a replacement $\partial_y \rightarrow -iq$ in (\ref{Killing vector})
leads to the angular momentum operator in the presence of a monopole with
magnetic charge $q$ at the origin \cite{Wu:1976ge}:
\begin{eqnarray}
&&L_1^{(q)}=i(\sin\varphi\partial_{\theta}+
\cot\theta\cos\varphi\partial_{\varphi})-
q\frac{1\mp \cos\theta}{\sin\theta}\cos\varphi, \nonumber\\
&&L_2^{(q)}=i(-\cos\varphi\partial_{\theta}+
\cot\theta\sin\varphi\partial_{\varphi})-
q\frac{1\mp \cos\theta}{\sin\theta}\sin\varphi, \nonumber\\
&&L_3^{(q)}=-i\partial_{\varphi}\mp q ,
\label{monopole angular momentum}
\end{eqnarray}
where $q$ is quantized as 
$q=0, \pm \frac{1}{2}, \pm 1, \pm \frac{3}{2},\cdots$,
because $y$ is a periodic variable with the period $4\pi$.
These operators act on the local sections on $S^2$ and satisfy the $SU(2)$
algebra $[L_i^{(q)},L_j^{(q)}]=i\epsilon_{ijk}L_k^{(q)}$.
When $q=0$, these operators are reduced to the ordinary angular momentum operators
 on $S^2$ (or $R^3)$,  $L_i\equiv L_i^{(0)}$. 
The $SU(2)$ acting on $g$ from left survives as the isometry of $S^2$.

\section{Proof of equation (\ref{zeta identity 2})}
\label{Proof of identity}

General `tree' planar diagram shown in 
Fig.\ref{tree planar diagram}
consists of several parts (blobs)
which are connected by the dashed lines and
each part represents a planar diagram.
For each part, we define a certain quantity in 
order to show (\ref{zeta identity 2}).
For instance, if we consider the left diagram shown in 
Fig.\ref{typical diagram in mm},
we define the following quantity for the part labeled by $s$,
\begin{align}
Q_s=
\sum_{\substack{t=-\Lambda/2 \\ (t\neq s)}}^{\Lambda /2}
\sum_{\substack{u=-\Lambda/2 \\ (u\neq t)}}^{\Lambda /2}
\sum_{\substack{v=-\Lambda/2 \\ (v\neq u)}}^{\Lambda /2}
\sum_{\substack{w=-\Lambda/2 \\ (w\neq u)}}^{\Lambda /2}
\frac{1}{(s-t)^{8}}
\frac{1}{(t-u)^{10}}
\frac{1}{(u-v)^{4}}
\frac{1}{(u-w)^{4}}.
\end{align}
Namely, for a dashed line connecting $s$ and $t$, 
we assign $\frac{1}{(s-t)^{2l}}$ where $2l$ is the number of
fields contained in the vertex and
$Q_s$ is the summation of the product of them with respect to the
variables other than $s$.
Then, we can see that $Q_s$ has a finite value
and bounded from above by a certain constant.
This is shown as follows.
First, note the following identity,
\begin{align}
\sum_{\substack{t=-\Lambda/2 \\ (t\neq s)}}^{\Lambda /2}
\frac{1}{(s-t)^{2l}}=
2\zeta(2l)-\zeta(2l,\Lambda/2+s+1)-\zeta(2l,\Lambda/2-s+1),
\end{align}
where $\zeta(z)$ and $\zeta(z,a)$ are the zeta function and the
generalized zeta function, respectively. They are defined by,
\begin{align}
\zeta (z)=\sum_{n=1}^{\infty}\frac{1}{n^z}, \;\;\;\;
\zeta (z, a)=\sum_{n=0}^{\infty}\frac{1}{(n+a)^z}.
\label{zeta function}
\end{align}
Because $\zeta(2l,\Lambda/2+s+1)+\zeta(2l,\Lambda/2-s+1)>0$,
\begin{align}
Q_s &<
\sum_{\substack{t=-\Lambda/2 \\ (t\neq s)}}^{\Lambda /2}
\sum_{\substack{u=-\Lambda/2 \\ (u\neq t)}}^{\Lambda /2}
\sum_{\substack{v=-\Lambda/2 \\ (v\neq u)}}^{\Lambda /2}
\frac{1}{(s-t)^{8}}
\frac{1}{(t-u)^{10}}
\frac{1}{(u-v)^{4}}
\times 2\zeta(4)
\nonumber\\
&< 
\sum_{\substack{t=-\Lambda/2 \\ (t\neq s)}}^{\Lambda /2}
\sum_{\substack{u=-\Lambda/2 \\ (u\neq t)}}^{\Lambda /2}
\frac{1}{(s-t)^{8}}
\frac{1}{(t-u)^{10}}
\times 2^2\zeta(4)^2
\nonumber\\
&< 
\sum_{\substack{t=-\Lambda/2 \\ (t\neq s)}}^{\Lambda /2}
\frac{1}{(s-t)^{8}}
\times 2^3\zeta(4)^2\zeta(10)
\nonumber\\
&< 2^4\zeta(4)^2\zeta(8)\zeta(10).
\end{align}
Thus, we find that $Q_s$ is finite even if we take the limit
$\Lambda \rightarrow \infty$.
We are able to define $Q_s$ also for more general diagram shown 
in Fig.\ref{tree planar diagram}.
If the diagram is `tree', through the same calculation 
just as we described above, we see that $Q_s$ is bounded 
from above by a $\Lambda$-independent constant which 
is given by a product of the zeta functions\footnote{Actually,
the upper bound of $Q_s$ exists also for non-`tree' diagrams.
However, this fact is not important for our purpose.}.

\begin{figure}[tbp]
\begin{center}
\includegraphics[height=3cm, keepaspectratio, clip]{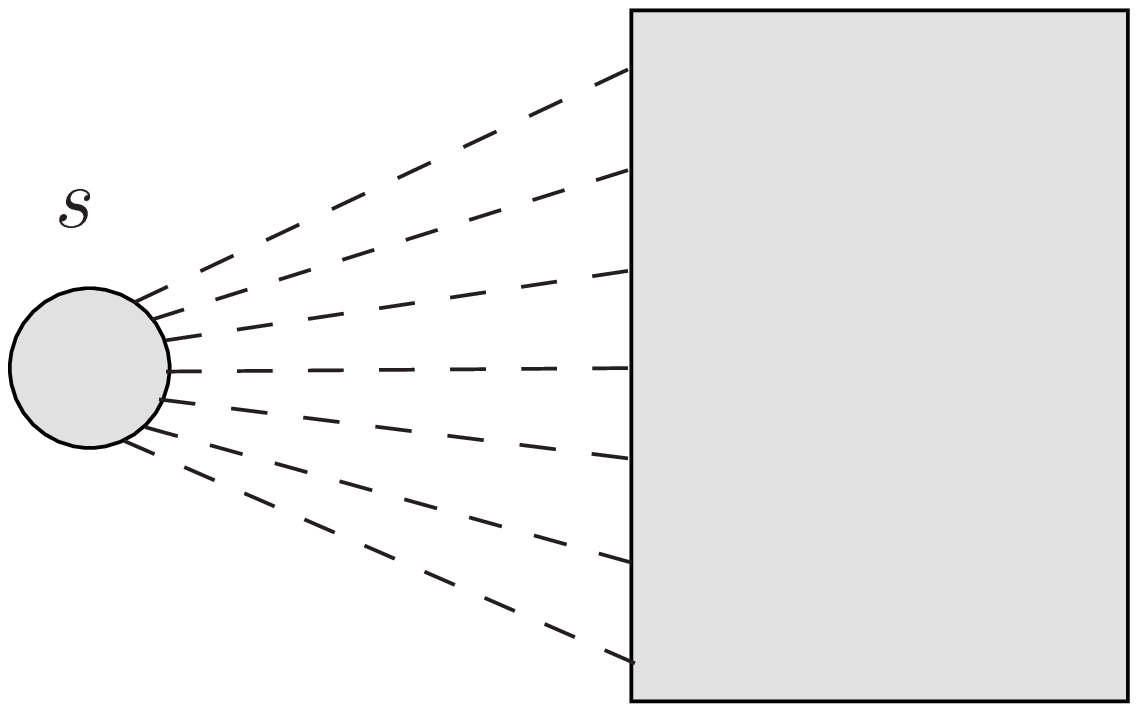}
\end{center}
\caption{The rectangular box represents a Feynman diagram 
with `tree'-like structure, namely, a diagram 
which does not contain any loop formed by dashed lines.
The circular blob represents a planar diagram which does
not include dashed lines.}
\label{Qs}
\end{figure}

Then, let us consider the diagram in Fig.\ref{Qs}.
We can define $Q_s$ for this diagram.
We assume that the rectangular part in Fig.\ref{Qs}
represents a `tree' diagram so that $Q_s$ for this diagram
is bounded from above.
We consider the following quantity,
\begin{align}
\frac{1}{\Lambda+1}\sum_{s=-\Lambda/2}^{\Lambda/2}
\sum_{\substack{t=-\Lambda/2 \\ (t\neq s)}}^{\Lambda /2}
\frac{1}{(s-t)^{2l}}Q_s
=\frac{2}{\Lambda+1}\sum_{s=-\Lambda/2}^{\Lambda/2}
(\zeta(2l)-\zeta(2l,\Lambda/2+s+1))Q_s.
\label{Qs1}
\end{align}
Since
there exists a positive $\Lambda$-independent
constant $c$ such that
$Q_s < c$, the second term 
on the right hand side in (\ref{Qs1}) 
is bounded as follows,
\begin{align}
\frac{1}{\Lambda+1}\sum_{s=-\Lambda/2}^{\Lambda/2}
\zeta(2l,\Lambda/2+s+1)Q_s
&<
\frac{c}{\Lambda+1}\sum_{s=-\Lambda/2}^{\Lambda/2}
\zeta(2l,\Lambda/2+s+1)
\nonumber\\
&= \frac{2c}{\Lambda+1}
\sum_{n=1}^{\Lambda+1}\frac{1}{n^{2l-1}}
+2c \sum_{n=\Lambda+1}^{\infty}
\frac{1}{n^{2l}},
\label{bound}
\end{align}
where we have used (\ref{zeta function}) to obtain
the second line.
Because the second line in (\ref{bound}) is 
${\cal O}(\log \Lambda /\Lambda)$ for $l=1$ and
${\cal O}(1 /\Lambda)$ for $l>1$, this quantity 
goes to zero in the limit $\Lambda \rightarrow \infty$.
Therefore, we find from (\ref{Qs1}) and (\ref{bound})
that 
\begin{align}
\lim_{\Lambda \rightarrow \infty}
\frac{1}{\Lambda+1}\sum_{s=-\Lambda/2}^{\Lambda/2}
\sum_{\substack{t=-\Lambda/2 \\ (t\neq s)}}^{\Lambda /2}
\frac{1}{(s-t)^{2l}}Q_s
=
\lim_{\Lambda \rightarrow \infty}
\frac{2\zeta(2l)}{\Lambda+1}
\sum_{s=-\Lambda/2}^{\Lambda/2}Q_s.
\label{formula 1}
\end{align}
Then, for instance,
we can show the following equality
by using (\ref{formula 1}) iteratively, 
\begin{align}
&\lim_{\Lambda\rightarrow\infty} \frac{1}{\Lambda+1}
\sum_{s=-\Lambda/2}^{\Lambda/2}
\sum_{\substack{t=-\Lambda/2 \\ (t\neq s)}}^{\Lambda /2}
\sum_{\substack{u=-\Lambda/2 \\ (u\neq t)}}^{\Lambda /2}
\sum_{\substack{v=-\Lambda/2 \\ (v\neq u)}}^{\Lambda /2}
\sum_{\substack{w=-\Lambda/2 \\ (w\neq u)}}^{\Lambda /2}
\frac{1}{(s-t)^{8}}
\frac{1}{(t-u)^{10}}
\frac{1}{(u-v)^{4}}
\frac{1}{(u-w)^{4}}
\nonumber\\
&=
\lim_{\Lambda\rightarrow\infty} \frac{1}{\Lambda+1}
\sum_{s=-\Lambda/2}^{\Lambda/2}
\sum_{\substack{t=-\Lambda/2 \\ (t\neq s)}}^{\Lambda /2}
\sum_{\substack{u=-\Lambda/2 \\ (u\neq t)}}^{\Lambda /2}
\sum_{\substack{v=-\Lambda/2 \\ (v\neq u)}}^{\Lambda /2}
\frac{1}{(s-t)^{8}}
\frac{1}{(t-u)^{10}}
\frac{1}{(u-v)^{4}}
\times 2\zeta(4)
\nonumber\\
&=
\lim_{\Lambda\rightarrow\infty} \frac{1}{\Lambda+1}
\sum_{s=-\Lambda/2}^{\Lambda/2}
\sum_{\substack{t=-\Lambda/2 \\ (t\neq s)}}^{\Lambda /2}
\sum_{\substack{u=-\Lambda/2 \\ (u\neq t)}}^{\Lambda /2}
\frac{1}{(s-t)^{8}}
\frac{1}{(t-u)^{10}}
\times 2^2\zeta(4)^2
\nonumber\\
& \;\;\; \vdots
\nonumber\\
&=\lim_{\Lambda\rightarrow\infty} \frac{1}{\Lambda+1}
\sum_{s=-\Lambda/2}^{\Lambda/2}
2^4\zeta(4)^2\zeta(8)\zeta(10)
\nonumber\\
&=2^4\zeta(4)^2\zeta(8)\zeta(10).
\label{proof}
\end{align}
This is nothing but (\ref{zeta identity}).
Since (\ref{proof}) implies that we can replace
$\sum_{s=-\Lambda/2}^{\Lambda/2}1/(s-t)^{2l}$ 
on the tip of a branch in the `tree' diagrams
with $\zeta(2l)$ in the $\Lambda \rightarrow \infty$ 
limit, applying the same calculation,
we can show (\ref{zeta identity 2}) for 
the generic `tree' planar diagram
shown in Fig.\ref{tree planar diagram}.

\section{Perturbative calculation}
\label{Perturbative calculation}

In this appendix, we explicitly evaluate the free energy of (\ref{mm covariant}) and the VEV of
(\ref{WL}) up to $\mathcal{O}(\lambda^3)$. 
We perform the usual perturbation theory in (\ref{mm covariant}). 

\begin{figure}[t]
\begin{center}
\includegraphics[scale=0.7]{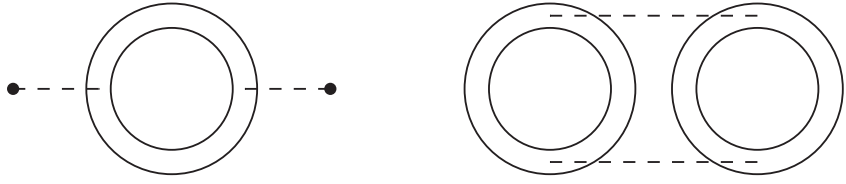}
\caption{Feynman diagrams corresponding to (\ref{FE ex}).
The left is planar diagram ($\mathcal{O}(N^2)$), while the right is nonplanar ($\mathcal{O}(1)$).}
\label{FE_example}
\end{center}
\end{figure}

First, we will calculate the free energy. 
Let us expand the interaction terms in terms of the power of the coupling
\begin{align}
\tilde{V}(\phi_s)&= \sum_{k=1}^{\infty}\tilde{V}_{k}(\phi_s), \notag\\
\tilde{V}_k(\phi_s)
&\equiv \sum_{s\neq t}
\frac{1}{2k(s-t)^{2k}}
\left(\frac{-\lambda}{4\pi^2}\right)^k 
\sum_{m=0}^{2k}{}_{2k}C_m {\rm tr} \phi_s^m
{\rm tr}(- \phi_t)^{2k-m},
\end{align}
and the partition function as
\begin{align}
\mathcal{Z}=\sum_{k=0}^{\infty}\mathcal{Z}_k,
\end{align}
where $\mathcal{Z}_k$ is the part of which the coupling dependence is $\mathcal{O}(\lambda^k)$.
$\mathcal{Z}_0$ is the free part,
\begin{align}
\mathcal{Z}_0
=\int \prod_s d\phi_s e^{-\frac{N}{2}\sum_s \tr \phi_s^2}
=\left(\frac{2\pi}{N}\right)^{N^2(\Lambda+1)/2},
\end{align}
and $\mathcal{Z}_k\; (k=1,2,3)$ are given by
\begin{align}
\frac{\mathcal{Z}_1}{\calZ_0}
&=-\langle \tilde{V}_1 \rangle, \notag\\
\frac{\mathcal{Z}_2}{\calZ_0}
&=-\langle \tilde{V}_2 \rangle+\frac{1}{2}\langle \tilde{V_1}^2 \rangle, \notag\\
\frac{\mathcal{Z}_3}{\calZ_0}
&=-\langle \tilde{V}_3 \rangle
+\langle \tilde{V_2}\tilde{V_1} \rangle -\frac{1}{3!}\langle \tilde{V_1}^3 \rangle.
\label{cZ_1,2,3}
\end{align} 
We define the free energy of our matrix model as
\begin{align}
\mathcal{F}
&=\ln \frac{\mathcal{Z}}{\mathcal{Z}_0} \nonumber\\
&=\ln \left(1+\frac{\calZ_1}{\calZ_0}+\frac{\calZ_2}{\calZ_0}+\frac{\calZ_3}{\calZ_0}
+\cdots \right).
\label{Free energy}
\end{align}
Substituting (\ref{cZ_1,2,3}) into (\ref{Free energy}), we obtain
\begin{align}
\mathcal{F}
=-\av{\tilde{V}_1}_c-\av{\tilde{V}_2}_c+\frac{1}{2}\av{\tilde{V_1}^2}_c-\av{\tilde{V}_3}_c
+\av{\tilde{V_2}\tilde{V_1}}_c-\frac{1}{3!}\av{\tilde{V_1}^3}_c+\cdots,
\end{align}
where $\av{\cdots}_c$ means the connected part of $\av{\cdots}$.
For example, $\av{\tilde{V}_1^2}_c$ is calculated as follows;
\begin{align}
\frac{1}{2}\av{\tilde{V_1}^2}_c
&=\frac{1}{2}\biggl(4 \sum_{s,t,u}\frac{1}{2(s-t)^2\cdot 2(t-u)^2}
\left(\frac{\lambda}{4\pi^2}\right)^2
N^2\av{\tr\phi^2\tr\phi^2}_c \n
&\qquad
+2 \sum_{s,t} \frac{1}{2(s-t)^2\cdot 2(t-s)^2}\left(\frac{\lambda}{4\pi^2}\right)^2
2^2 \av{\tr\phi\tr\phi}_c^2
\biggr) \n
&=\frac{\lambda^2}{4\pi^4}N^2(\Lambda+1)\zeta(2)^2
+\frac{\lambda^2}{8\pi^4}(\Lambda+1)\zeta(4) \n
&=(\Lambda+1)\left( N^2 \frac{\lambda^2}{144}+\frac{\lambda^2}{720} \right).
\label{FE ex}
\end{align}
In Feynman diagram introduced in subsection 
\ref{Feynman rule for our matrix model},
 one can describe this contribution  as Fig. \ref{FE_example}.
Performing the similar calculations, we finally obtain
(\ref{pc 1}).

\begin{figure}[t]
\begin{center}
\includegraphics[scale=0.5]{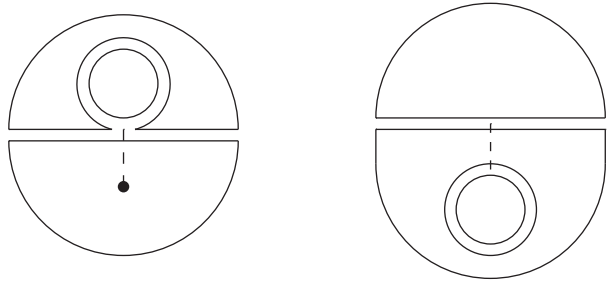}
\caption{Wilson loop corresponding to the leading part of (\ref{WL ex}).} \label{WL_example}
\end{center}
\end{figure}

Second, we perform a perturbative calculation of the VEV of the unknot Wilson loop (\ref{WL})
up to $\mathcal{O}(\lambda^3)$.
We expand the VEV as 
\begin{align}
\frac{1}{(\Lambda+1)N}\sum_s \av{\Tr e^{\sqrt{\lambda}\phi_s}}
&=\frac{1}{(\Lambda+1)N} \sum_{l=0}^{\infty}\frac{\lambda^{l/2}}{l!}\sum_s\av{\tr(\phi_s)^l} \n
&=1+\frac{1}{(\Lambda+1)N}
 \sum_{l=1}^{\infty}\frac{\lambda^{l}}{(2l)!}
 \sum_s(\av{\tr(\phi_s)^{2l}}_{0}+\av{\tr(\phi_s)^{2l}}_{int}), \label{pert WL}
\end{align}
where $\av{\cdots}_0$ means the ladder part and $\av{\cdots}_{int}$ means to evaluate 
with interaction vertices.
The ladder part is easily calculated as follows
\begin{align}
\sum_s\av{\tr\phi_s^2}_{0}&=(\Lambda+1)N,\n
\sum_s\av{\tr\phi_s^4}_{0}&=(\Lambda+1)\left(2N+\frac{1}{N}\right),\n
\sum_s\av{\tr\phi_s^6}_{0}&=(\Lambda+1)\left(5N+\frac{10}{N}\right).
\end{align}
Then, up to $\mathcal{O}(\lambda^3)$ the ladder part gives
\begin{align}
\frac{1}{(\Lambda+1)N}\sum_{l=0}^{\infty}\frac{\lambda^{l}}{(2l)!}\sum_{s}\av{\phi_s^{2l}}_{0}
&=1+\frac{\lambda}{2}+\frac{\lambda^2}{4!}\left(2+\frac{1}{N^2}\right)
+\frac{\lambda^3}{6!}\left(5+\frac{10}{N^2}\right)+\cdots. \label{pert WL ladder}
\end{align}
Next, let us expand the third term in (\ref{pert WL}) up to $\mathcal{O}(\lambda^3)$
\begin{align}
&\frac{1}{(\Lambda+1)N}
 \sum_{l=1}^{\infty}\frac{\lambda^{l}}{(2l)!}
 \sum_s \av{\tr(\phi_s)^{2l}}_{int} \n
&=\frac{1}{(\Lambda+1)N}
 \biggl[\frac{\lambda}{2!}
 \sum_s
 \left(-\av{\tr(\phi_s)^2\tilde{V}_1}_c-\av{\tr(\phi_s)^2\tilde{V}_2}_c
 +\frac{1}{2}\av{\tr(\phi_s)^2\tilde{V}_1^2}_c
\right)
\nonumber\\
& \hspace{7.5cm} -\frac{\lambda^2}{4!}
\sum_s \av{\tr(\phi_s)^4\tilde{V}_1}_c+\cdots
\biggr]. \label{pert WL int}
\end{align}
For example, we can calculate the second term in (\ref{pert WL int}) as
\begin{align}
-\sum_s\av{\tr\phi_s^2\tilde{V}_2}_c
&=-\frac{\lambda^2}{32\pi^4}
\left(
2\times N\av{\tr\phi^2\tr\phi^4}_c+6\av{\tr\phi^2\tr\phi^2}_c\av{\tr\phi^2}_c\times 2
\right)
\sum_{s\neq t} \frac{1}{2(s-t)^{4}} \n
&=-\frac{5\lambda^2N}{4\pi^4}
\left(
1+\frac{1}{5N^2}
\right)
 (\Lambda+1)\zeta(4) \n
&=-(\Lambda+1)N\left( 1+\frac{1}{5N^2} \right)  \frac{\lambda^2}{72}. \label{WL ex}
\end{align}
The Feynman diagrams corresponding to the leading part are depicted in Fig. \ref{WL_example}.
We can calculate the other terms in the same way and the result is
\begin{align}
&\frac{1}{(\Lambda+1)N}\sum_{l=1}^{\infty}\frac{\lambda^{l}}{(2l)!}\sum_{s}\av{\phi_s^{2l}}_{int} \n
&=
\frac{\lambda}{2}\left(
\frac{\lambda}{6} -\left( 1+\frac{1}{5N^2} \right)  \frac{\lambda^2}{72}
+\frac{\lambda^2}{36}+\frac{1}{N^2}\frac{\lambda^2}{180}
+\cdots
\right)
+\frac{\lambda^2}{4!}\left(1+\frac{1}{2N^2}\right)\frac{2\lambda}{3} +\cdots.
\label{pert WL int 2}
\end{align}
Finally, gathering (\ref{pert WL ladder}) and (\ref{pert WL int 2}), 
we obtain perturbative expansion 
of the unknot Wilson loop in our matrix model
(\ref{VEV of Wilson loop in matrix model}).

\end{document}